\documentclass[aps,prb,twocolumn,floats,epsfig]{revtex4-2}
\usepackage[iso-8859-1]{inputenx}
\usepackage{amssymb}
\usepackage{amsbsy}
\usepackage{amsmath}
\usepackage{epsfig}
\usepackage{color}
\usepackage{float}
\usepackage[colorlinks,linktocpage,bookmarks=false,citecolor=blue,linkcolor=red,urlcolor=blue]{hyperref}

\newcommand{\beq}{\begin{equation}}
\newcommand{\eeq}{\end{equation}}
\newcommand{\bea}{\begin{eqnarray}}
\newcommand{\eea}{\end{eqnarray}}

\newcommand{\si}{\sigma}

\newcommand{\om}{\omega}

\newcommand{\non}{\nonumber}
\newcommand{\noi}{\noindent}

\usepackage{xcolor}

\begin{document}

\title{Quantum dynamics of a spin model with an extensive degeneracy}

\author{Krishanu Ghosh$^1$, Diptiman Sen$^2$ and K. Sengupta$^1$}

\affiliation{$^{(1)}$School of Physical Sciences, Indian Association for the
Cultivation of Science, Jadavpur, Kolkata 700032, India \\
$^{(2)}$Center for High Energy Physics, Indian Institute of Science, Bengaluru 560012, India}

\date {\today}


\begin{abstract}

We study the role played by extensive degeneracy in shaping the nature of the quantum dynamics of a one-dimensional spin model for both ramp and periodic drive protocols. The model displays an extensive degenerate manifold of states for a specific value of one of the parameters of its Hamiltonian. We study a linear ramp which takes the spin model through this degenerate point and show that it leads to a deviation from the usual Kibble-Zurek behavior. We also  study the St\"uckelberg oscillations in such a model for a ramp which passes twice through the degenerate point. Our study indicates that such oscillations are strongly suppressed leading to a distinct behavior compared to those 
arising from double passage through a quantum critical point. Finally, we study 
the periodic dynamics of the model and show, for a large drive amplitude, the existence of special drive frequencies at which the system exhibits an approximate emergent $U(1)$ 
symmetry. We study the effect of this emergent symmetry on the correlators of the driven system and demonstrate the existence of 
dynamic symmetry restoration at these 
frequencies. We study the fate of the emergent symmetry when the drive amplitude is 
decreased and discuss possible experiments to test our theory.

\end{abstract}

\maketitle 

\section{Introduction}
\label{secint}

Degeneracies are an interesting feature of generic quantum systems. The presence of an extensive degeneracy in 
interacting quantum systems is key to understanding the physics of, for example, fractional quantum Hall systems \cite{frrev1} and 
spin liquids \cite{slrev1}. For fractional quantum Hall systems, a strong magnetic field in the kinetic term of the Hamiltonian of non-interacting 
electrons leads to an extensive degeneracy; the mechanism of the lifting of this degeneracy by Coulomb interactions between the electrons is central to the realization of
fractional quantum Hall states hosting quasiparticle excitations with fractional charge and anyonic statistics \cite{laughlin1,jain1}. For typical spin liquids, the relevant degeneracy can be often be understood in terms of an
effective electrodynamics with the degenerate manifold being formed by all states satisfying the corresponding Gauss law \cite{slrev1,slrev2}. The 
presence of such degeneracy-induced phenomena provides a motivation for studying them in simpler contexts where their {\color {blue} features}
can be understood analytically \cite{obdref1,obdref2,as1}. 

The non-equilibrium dynamics of interacting closed quantum systems has been studied extensively in recent years \cite{rev1,rev2,rev3,rev4,rev5,rev6,rev7,rev8,rev9,rev10,rev11,rev12}. The initial endeavors in this direction focused on quench 
and ramp dynamics in such systems \cite{rev1,rev2,rev3,rev4}. Such ramps through quantum critical points of these quantum models 
have been of particular interest; it is well known that for slow ramps they lead to Kibble-Zurek scaling for the fidelity and 
residual energies \cite{kz1,kz2,ap1,ds1,ds2}. Several studies have also focused on ramps which lead to multiple passages through 
such critical points; these studies show the existence of St\"uckelberg oscillations leading to an oscillatory behavior of, for example, 
residual energies as a function of the ramp rate \cite{rev4,bhaskar1}. 

In more recent years the study of non-equilibrium dynamics of periodically driven systems have received a lot of attention \cite{rev5,rev6,rev7,rev8,rev9,rev10,rev11,rev12}. These systems are typically described by an evolution operator, which, at stroboscopic times 
$t= mT$ ($m \in Z$), can be described by the Floquet Hamiltonian $H_F$: $U(mT,0)= \exp[-i H_F(T) mT]$. Here $T=2 \pi/\omega_D$ 
denotes the drive period and $\omega_D$ is the drive frequency. The computation of $H_F$ therefore allows one to understand the 
stroboscopic dynamics of the driven system. However, an exact computation of $H_F$ is usually impossible for interacting quantum systems; this has led to several perturbative 
techniques for its computation, specially in the regime of high drive amplitude or frequency where $H_F$ 
has a local structure \cite{rev7}. Such driven systems are known to exhibit several phenomena such as dynamical freezing \cite{ dynfr1,dynfr2,dynfr3,dynfr4}, dynamical 
localization \cite{dloc1,dloc2,dloc3,dloc4,dloc5}, emergent topological features \cite{topo1,topo2,topo3,topo4}, Floquet realization of quantum scars \cite{qscar1,qscar2,qscar3,qscar4} and Hilbert space fragmentation \cite{som1,dsf}, and 
time crystalline states of matter \cite{tc1,tc2,tc3,tc4}. These phenomena have no analogs either in equilibrium or for systems subjected to aperiodic drives; many of 
these can be shown to be a consequence of approximate emergent symmetries of the Floquet Hamiltonian of the driven systems \cite{rev12}. 

In this work, we shall study the non-equilibrium dynamics for a one-dimensional (1D) spin model which has an extensive degeneracy for a special value of one of the parameters of its Hamiltonian. 
The model, discussed in detail in Sec.\ \ref{model}, hosts a large manifold of degenerate ground states at this special point; the number of these states scales 
exponentially with the system size $L$. In what follows, we study a ramp protocol which takes the system through this degenerate point
with a rate $\tau^{-1}$. Our analysis shows that the behavior of the system with such an extensive degeneracy is qualitatively different from their counterparts
for ramps through a critical point or for gapped systems. We compute the wave function overlap $F(t)$ and residual energy $Q(t)$ of such a system after a single passage 
through this degenerate point as a function of $t/\tau$ and show that they bear signatures of this degeneracy. We also study a ramp protocol which takes the system 
twice through such a degenerate point; we find that the presence of such extensive degeneracy leads to a strong suppression of St\"uckelberg oscillations \cite{rev4,bhaskar1} in $Q(t)$ at low ramp rates. 

We will also study the properties of the spin model in the presence of a periodic drive. We provide an analytic, albeit perturbative, computation of the Floquet Hamiltonian of 
the model for a square pulse protocol. Our analysis shows that in the high drive amplitude regime, the perturbative Floquet Hamiltonian harbors an approximate emergent 
$U(1)$ symmetry at special drive frequencies. We identify these frequencies and show that for high drive amplitude and at the special frequencies, the behavior for certain correlation functions of the model is controlled by the emergent symmetry up to a long prethermal timescale. We also study the behavior of such correlation 
functions at these drive frequencies starting from initial spin states which break the emergent symmetry; we find that the emergent symmetry is dynamically restored at long 
times. This provides an example of dynamical symmetry restoration studied recently in the context of other models \cite{cala1, cala2, tista1}. 

The plan of the rest of this paper is as follows. In Sec.\ \ref{model}, we define the spin model and demonstrate that it exhibits an extensive degeneracy at a special parameter 
value. This is followed by Sec.\ \ref{rampd} where we discuss the ramp dynamics of the followed. Next, in Sec.\ \ref{periodd}, we 
analyze the periodic dynamics of the model and chart out the role of the emergent $U(1)$ symmetry in shaping the dynamics of its spin correlators. Finally, we discuss our main 
results, point out possible experiments that can verify them, and conclude in Sec.\ \ref{diss}. The details of the
calculation of the Floquet Hamiltonian are charted out in the Appendix.

\section{ Model and Degeneracy}
\label{model}

The Hamiltonian of the 1D spin model that we study is given by $H= H_0+H_1$ where
\begin{eqnarray} 
H_0 &=& -h \sum_j \sigma_j^x, \nonumber\\
H_1 &=& \frac{V_0}{4} \sum_{j} (1+\sigma_j^z)(1+\sigma_{j+1}^z).
\label{hamsys}
\end{eqnarray}
Here $\sigma_j^z$ and $\sigma_j^x$ are standard Pauli matrices, $j$ denotes the site index, $h$ is a transverse field, and $V_0$ denotes the strength of the coupling 
between neighboring spins. We note here that this model constitutes an effective description of a Rydberg chain at zero detuning
and in the limit where the Van der Waals interaction is small enough to be approximated by the nearest-neighbor term \cite{rydexp1,rydexp2,rydexp3,rydexp4}. In the context of the Rydberg chain, 
the Pauli matrix $\sigma_j^z$ is related to the density of Rydberg excitations 
$\hat n_j$ by $ \hat n_j = (1+\sigma_j^z)/2$, while $\sigma_j^x = 
|G_j\rangle \langle R_j| + |R_j\rangle \langle G_j|$, where $|G_j\rangle$ and 
$|R_j\rangle$ denote the ground and Rydberg excited states respectively
of an atom at site $j$.

\begin{figure}
\includegraphics[width=0.98\linewidth]{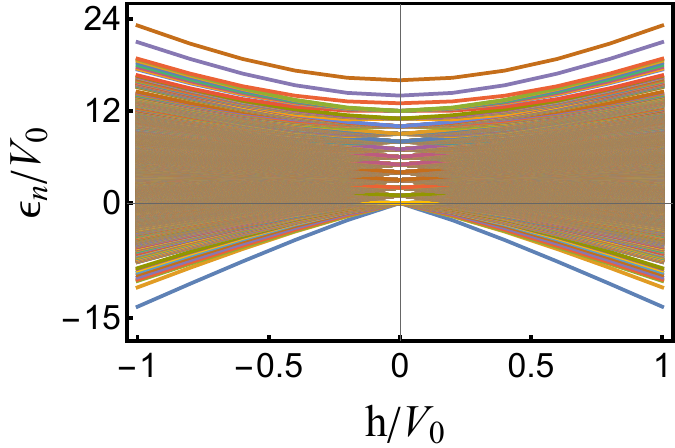}
\caption{Plot of energy labels $\epsilon_n/V_0$ of $H$ (Eq.\ \eqref{hamsys}) as a 
function of $h/V_0$ indicating an extensive degeneracy at $h =0$. 
The system size $L=16$ and $V_0=1$.} \label{figm1} \end{figure}

The extensive degeneracy of eigenstates of $H_1$ (Eq.\ \eqref{hamsys}), and hence of the eigenstates of $H$ at $h=0$, can be inferred as follows. The degenerate manifold of 
states corresponds to Fock states where no two neighboring spins at sites $j$ and $j+1$ are in the $|\uparrow\rangle$ state. The number of states obeying such a 
constraint for a chain of length $L$ (in units of lattice spacing $a$ which is set to unity in the rest of this work) with periodic boundary
conditions can be exactly computed using a straightforward transfer 
matrix method, and is given by \cite{qscar1,qscar2} 
\begin{eqnarray} {\mathcal N}(L) &=& \varphi^L + (-1/\varphi)^L, \label{statecount} 
\end{eqnarray} 
where $\varphi= (1+\sqrt{5})/2$ is the golden ratio. For large $L$, this leads to an exponentially large number ($N_L \simeq \varphi^L$) of degenerate states. This is shown in 
Fig.\ \ref{figm1}, where the energy eigenvalues $\epsilon_n$ of $H$ are plotted as a 
function of $h$. The plot shows an extensive manifold of degenerate zero energy 
eigenstates for $h=0$. In what follows, we study the effect of such a degeneracy when the transverse field is tuned through the degeneracy point, using either a ramp (Sec.\ 
\ref{rampd}) or a periodic (Sec.\ \ref{periodd}) protocol. 

Before ending this section, we note that while studying the ramp dynamics of this model in the next section, we shall compare its behavior to that found in a model 
with a conventional quantum
critical point. For this purpose, we choose the well-known PXP model in a longitudinal field \cite{subir1,subir2,pekker1,qscar1,qscar2} whose Hamiltonian is given by
\begin{eqnarray}
H' &=&  \sum_j ~(-w \tilde \sigma_j^x +\lambda \sigma_j^z/2), \nonumber\\
\tilde \sigma_j^x &=& P_{j-1} \sigma_j^x P_{j+1}, \quad P_j=(1-\sigma_j^z)/2.\label{critham} \end{eqnarray}
Here $P_j$ is the projector to the spin $\downarrow$ state at site $j$ and 
$|\lambda|\gg w$ with $\lambda<0$ represents the $Z_2$ symmetry broken phase of the model \cite{subir1}.
It is well-known that the first term of $H'$ represents the Hamiltonian given by Eq.\ \eqref{hamsys} in the 
Rydberg blockade regime where the interaction term forbids the appearance of two neighboring up-spins, for $V_0/2^6 \ll w \ll V_0$ \cite{rydexp1,rydexp2,rydexp3,rydexp4}. The second term can be obtained from a 
finite detuning term for the Rydberg atoms \cite{rydexp1,rydexp2,rydexp3,rydexp4}. The model harbors a critical point at $\lambda=\lambda_c=-1.31 w$ which belongs to the Ising universality class. Both 
the equilibrium \cite{subir1,subir2} and non-equilibrium \cite{pekker1,qscar1,qscar2} properties of this model have been extensively studied in the literature.

\section{Ramp dynamics}
\label{rampd} 

In this section, we study the ramp dynamics of the spin system whose Hamiltonian is given by Eq.\ \eqref{hamsys}. To this end, we consider a linear ramp of the transverse field from $-h_0$ to $h_0$ 
with a rate $\tau^{-1}$ given by
\begin{eqnarray} h(t) ~=~ h_0 ~(\frac{2t}{\tau} ~-~ 1). \label{rampprot}
\end{eqnarray}
Note that the degenerate point is traversed at $t=\tau/2$. To study the dynamics of the system in the presence of such a ramp, we begin with an initial state $|\psi(0)\rangle$ 
and write down the Schrodinger equation $i \hbar \partial_t |\psi(t)\rangle= H(t) |\psi(t)\rangle$, where
\begin{eqnarray}
H(t) &=& H_f + \delta H(t), \nonumber\\
H_f &=& - ~h_0 ~\sum_j \sigma_j^x ~+~ \frac{V_0}{4} ~\sum_j (1+\sigma_j^z)(1+\sigma_{j+1}^z), \nonumber\\
\delta H(t) &=& 2h_0 ~(1-\frac{t}{\tau}) ~\sum_j \sigma_j^x. \label{hamramp}
\end{eqnarray} 
Next, we consider the eigenbasis of $H_f$; we denote the eigenstates as $|n\rangle$ and the corresponding eigenvalues as $\epsilon_n^f$. These can be computed numerically using 
exact diagonalization (ED) for a finite chain with length $L$. Expressing $|\psi(t)\rangle= \sum_n c_n(t) |n\rangle$, we find that
\begin{eqnarray}
i\hbar \frac{d c_n(t)}{d t} &=& \epsilon_n^f c_n(t) ~+~ \sum_m c_m(t) ~\Lambda_{mn}(t), \nonumber\\
\Lambda_{mn}(t) &=& \langle n| \delta H(t) |m\rangle. \label{rampeq}
\end{eqnarray}
In what follows, we will solve Eq.\ \eqref{rampeq} numerically with the initial condition $c_n(0)= \langle n|\psi(0)\rangle$ to obtain the state $|\psi(t)\rangle$. 

To study the behavior of the system under a ramp, we look at the fidelity $F(t)$ and the residual energy $Q(t)$ given by 
\begin{eqnarray}
F(t) &=& \ln |\langle\psi(t)|\psi_G(t)\rangle|^2, \nonumber\\
Q(t) &=& \langle \psi(t)|H(t)|\psi(t)\rangle ~-~ E_G(t), \label{fqeq} 
\end{eqnarray}
for several representative values of $\tau$. Here $|\psi_G(t)\rangle$ and $E_G(t)=\langle \psi_G(t)|H(t)|\psi_G(t)\rangle$ are the instantaneous ground state wave function and energy respectively.
It is well known that for a slow ramp which passes through a critical point with dynamical critical exponent $z$ and correlation length exponent $\nu$, $F(\tau) \sim \tau^{-d\nu/(z \nu+1)}$ and 
$Q(\tau) \sim \tau^{-(d+z)\nu/(z \nu+1)}$ for a $d$-dimensional system, in accordance with Kibble-Zurek scaling \cite{kz2,ap1}

\begin{figure}
\includegraphics[width=0.48\linewidth]{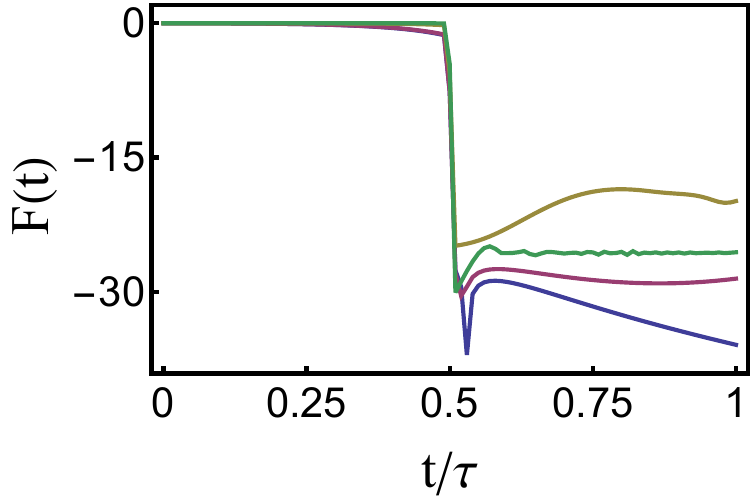}
\includegraphics[width=0.48\linewidth]{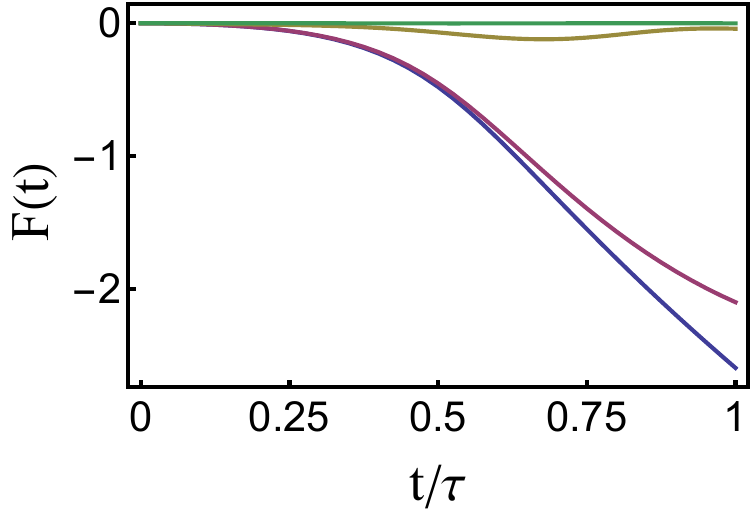}
\includegraphics[width=0.48\linewidth]{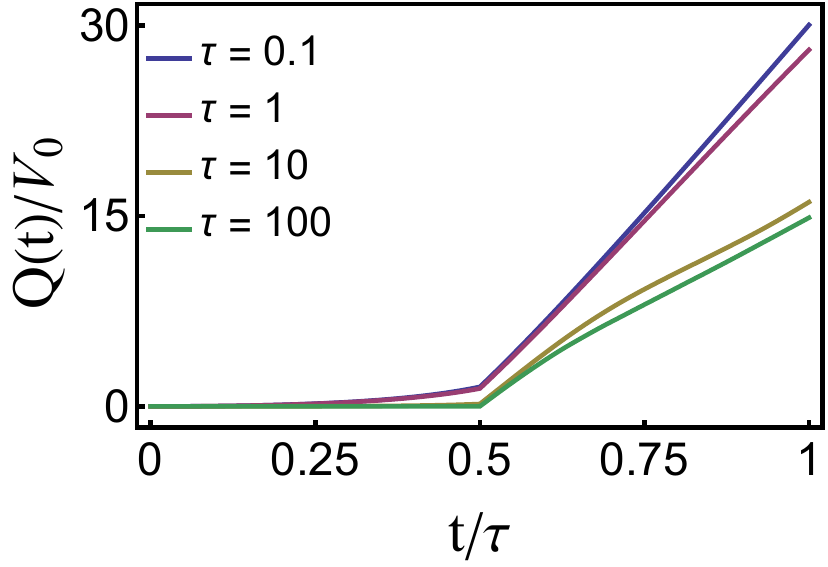}
\includegraphics[width=0.48\linewidth]{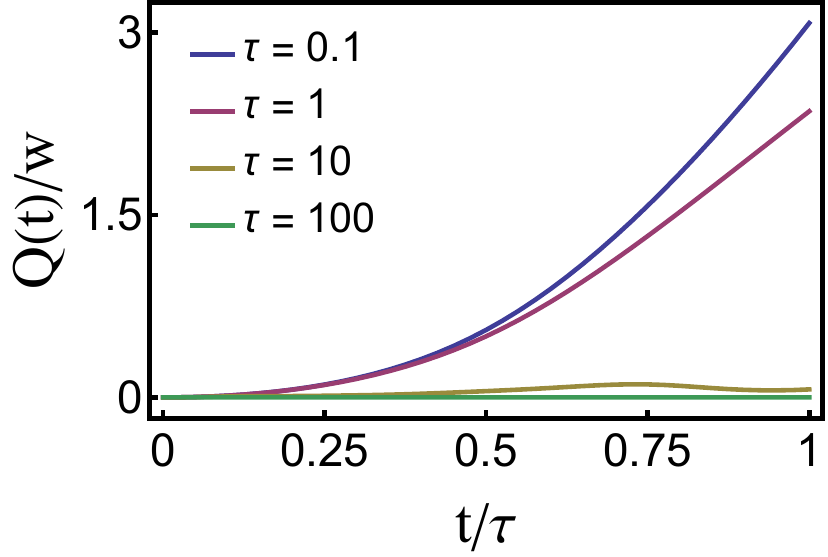}
\caption{Top left panel: Plot of $F(t)$ as a function of $t/\tau$ during a ramp from $t=0$ to $t=\tau$ 
for several representative values of $\tau$ and for the Hamiltonian $H$ (Eq.\ \eqref{hamsys}) which has an extensive degeneracy at $t/\tau = 1/2$. 
$V_0=1$ for all plots. Top right panel: Plot $F(t)$ as a function of $t/\tau$ for the Hamiltonian $H'$ (Eq.\ \eqref{critham}) where $\lambda$ is ramped 
through the critical point. For all plots $w=1=\lambda_0$. Bottom left panel: Plot of the residual energy $Q(t)/V_0$ as a function of $t/\tau$ for $H$. All parameters are the same 
as in the top left panel. Bottom right panel: 
Plot of $Q(t)/w$ as a function of $t/\tau$ for $H'$. All parameters are same as in the top right panel. For all figures $L=16$.} \label{figr1}
\end{figure}

We first study the behaviors of $F(t)$ and $Q(t)$ as a function of $t/\tau$ for several representative values of $\tau$ as shown in Fig.\ \ref{figr1}. The top left panel of Fig.\ \ref{figr1} shows the plot of $F(t)$ as a function of $t/\tau$ for a system with 
extensive degeneracy modeled by $H$ (Eq.\ \eqref{hamsys}), while the top right panel shows its behavior for a ramp of $\lambda$ through the critical point of $H'$ (Eq.\ \eqref{critham}). For the latter model, the ramp is given by
\begin{eqnarray} 
\lambda(t) &=& \lambda_c ~+~ \lambda_0 ~(\frac{2t}{\tau} -1), \label{rampcrit}
\end{eqnarray}
so that the critical point is traversed at $t=\tau/2$. 

The plot in the top left panel of Fig.\ \ref{figr1} clearly shows that upon passing through the degenerate point, the state of the system gains a finite weight on an 
extensive number of instantaneous eigenstates. Consequently $F(t)$ shows a sharp drop to large negative values for all $\tau$. In contrast, the top right panel of 
Fig.\ \ref{figr1} shows that $F(t)$ depends sensitively on $\tau$ for a ramp through the critical point. For slow drives parameterized by $\tau \gg 1$, $F(t)$ stays close to 
its initial value, while it decays to small values for faster ramp with $\tau \le 1$. A similar difference in behavior of $Q(t)$ can be seen by comparing the bottom panels. For 
a system with extensive degeneracy, $Q(t)$ increases linearly with $t$ for all $\tau$ once the degenerate point is traversed at $t=\tau/2$; in contrast, for a ramp through 
a critical point, its behavior depends on $\tau$. Moreover, $Q(\tau)$ attains a much larger value for a ramp through the degenerate point. 

This behavior of $Q(t)$ and $F(t)$ for a degenerate system can be qualitatively understood from Fig.\ \ref{figm1}. Fig.\ \ref{figm1} shows that an exponentially large 
number of instantaneous eigenstates of $H$ become degenerate for $h(t)=0$ at $t=\tau/2$. At this point, $|\psi(t)\rangle$ develops a finite overlap with these degenerate 
eigenstates with $|\langle \psi|\epsilon_n\rangle|^2 \sim 1/{\mathcal N}$ (Eq.\ \eqref{statecount}). Since ${\mathcal N} \sim e^L$ (Eq.\ \eqref{statecount}), these 
eigenstates have a typical gap $\Delta \epsilon \sim e^{-L}$. Consequently, $F$ shows a sharp drop for any $\tau > e^{-L}$. This situation is to be contrasted with that for 
a ramp through a critical point where only the ground state gap scales as $1/L^2$. In this case, for $\tau \ge 1$, the overlap of $|\psi(t)\rangle$ with the instantaneous eigenstates is significant only for the first few excited states, leading to a much 
larger value of $F(t)$ after passage through the critical point. The behavior of $Q(t)$ can also be understood similarly. The extensive spread of $|\psi(t)\rangle$ at the 
degenerate point leads to a fast growth $Q(t)$ as one moves away from this point during the ramp and leads to a much larger value of $Q(\tau)$. 

\begin{figure}
\includegraphics[width=0.48\linewidth]{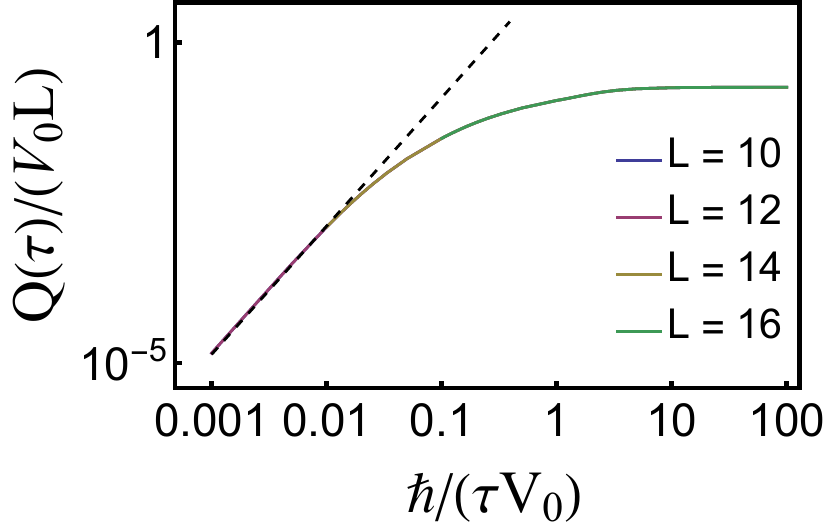}
\includegraphics[width=0.48\linewidth]{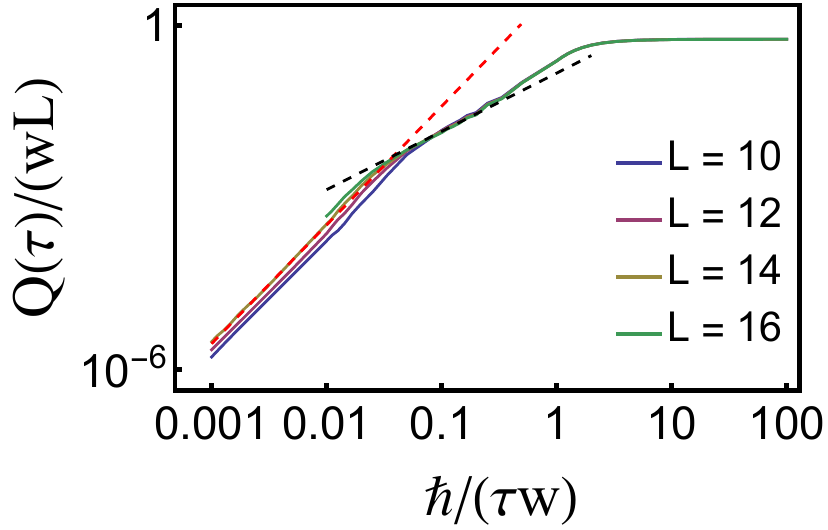}
\caption{Left panel: Plot of $Q(\tau)/(V_0 L)$ as a function of $\tau^{-1}$ for the Hamiltonian 
$H$ (Eq.\ \eqref{hamsys}) which hosts an extensive degeneracy. 
The dotted line shows a fit to $Q(\tau) \sim a/\tau^b$ with $a=14.01$ and $b=2.04$. For all plots $h_0/V_0=5$ $t_i=0$, $t_f=\tau/2$ 
and $V_0=1$. Right panel: Plot $Q(\tau)/(w L)$ as a 
function of $\tau$ for $H'$ (Eq.\ \eqref{critham}), where $\lambda$ is ramped 
to the critical point. For this plot, $t_i=0$, $t_F=\tau/2$, and $\lambda_0/w=5$. The red (black) dotted lines are fits to $Q(\tau) \sim a/\tau^b$ with $a=4.62 ~(0.14)$ and $b=2.07 ~(1.02)$. See text for details.} \label{figr2}
\end{figure}

Next we consider the behavior of $Q(\tau)$ as a function of $\tau$. It is well-known 
that for a finite system with length $L$ where the ramp takes one up to the critical point, 
$Q(\tau)$ displays either Kibble-Zurek scaling ($Q(\tau) \sim \tau^{-1}$ 
for quantum critical points with Ising universality) or Landau-Zener scaling ($Q(\tau) \sim \tau^{-2}$). The crossover between 
these two regimes occur when $\tau^{-1} \sim L^2$ for a critical point with Ising universality \cite{pekker1}. In contrast, for a ramp given by Eq.\ \eqref{rampprot} for $0\le t\le \tau/2$ which takes the system to the 
extensive degenerate point, $Q(\tau)$ is independent of $L$; moreover, it shows a direct crossover between the Landau-Zener scaling regime at large $\tau$ followed by a plateau at small $\tau$. Such a behavior follows from the presence of exponentially small instantaneous energy gaps at the degenerate point and leads to the absence of  Kibble-Zurek scaling for finite systems (shown for $L\le 16$ in the left panel of Fig.\ \ref{figr2}). This behavior is in sharp contrast to that found for a ramp through the critical point \cite{pekker1} which shows both Landau-Zener and Kibble-Zurek regimes for $L\le 16$ (shown 
in the right panel of Fig.\ \ref{figr2}).

\begin{figure}
\includegraphics[width=0.48\linewidth]{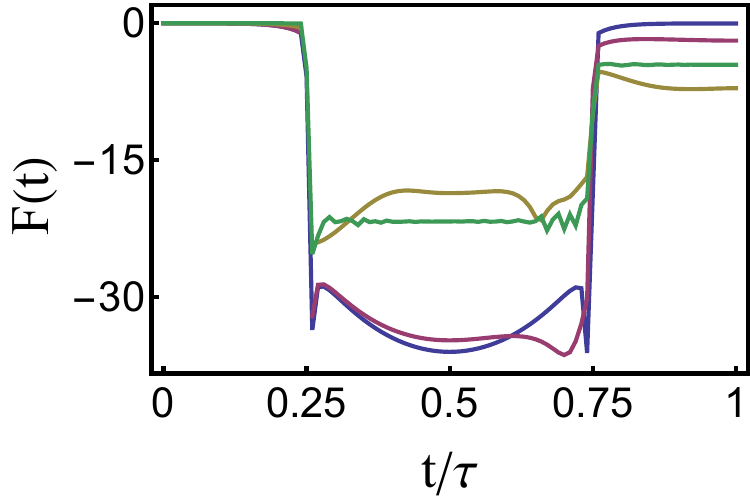}
\includegraphics[width=0.48\linewidth]{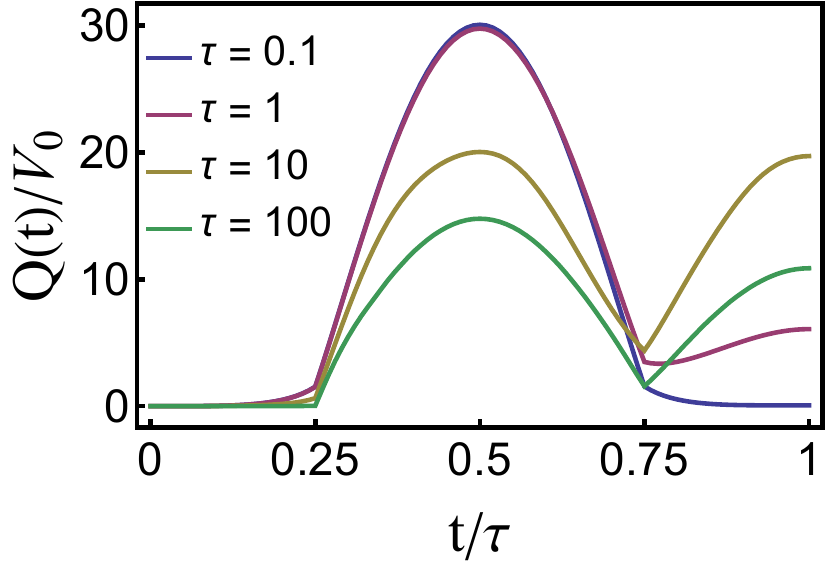}
\caption{Left panel: Plot of $F(t)$ as a function of $t/\tau$ where the system passes twice through the degenerate point at $t=\tau/4$ and $3\tau/4$ 
(Eq.\ \eqref{studr}) for several representative values of $\tau$ and for the Hamiltonian $H$ (Eq.\ \eqref{hamsys}) which hosts an extensive degeneracy. 
For all plots $V_0=1$. Bottom left panel: Plot of the residual energy $Q(t)/V_0$ for the same protocol. See text for details.} \label{figr3}
\end{figure}

Next, we consider a double passage through the degenerate point with a ramp given by
\begin{eqnarray} 
h(t) &=& h_0 ~\cos (2 \pi t/\tau), \quad 0\le t\le \tau, \label{studr} \end{eqnarray}
so that the degenerate point is crossed at $t=\tau/4$ and $3\tau/4$. 
The results for $F(t)$ and $Q(t)$ are shown in Fig.\ \ref{figr3}. We find that the second passage through the degenerate points leads to an
enhancement of $F$ and a reduction of $Q$. In contrast to the ramp studied earlier 
where the system passes through the degenerate point only once, these features depend sensitively 
on $\tau$. For smaller values of $\tau$, both $F(\tau)$ and $Q(\tau)$ are found
to attain values much closer to zero in Fig.~\ref{figr3} compared to the top and
bottom left panels of Fig.~\ref{figr1}. 
This can be understood as follows. We note that for smaller values of
$\tau$ mean that the wave function does not get much time to change and therefore
the wave functions at time $t=0$ and $\tau$ will be close to each other:
$| \psi (\tau) \rangle \simeq | \psi (0) \rangle = | \psi_G (0) \rangle$. For a single 
passage through the degenerate point (Eq.~\eqref{rampprot}),
the initial and final Hamiltonian are quite different from each other since
$h = - h_0$ and $h_0$ at $t=0$ and $\tau$ respectively. As a result,
the ground states of the initial and final Hamiltonian are quite different from
each other. Hence, the overlap $\langle \psi (\tau) | \psi_G (\tau) \rangle
\simeq \langle \psi_G (0) | \psi_G (\tau) \rangle$ will be 
small; hence $F(\tau)$ will be highly negative and $Q(\tau)$ will be large, as
we see in the left panels of Fig.~\ref{figr1}. In contrast,
for a double passage through the degenerate point (Eq.~\eqref{studr}), the initial
and final Hamiltonian are identical as both have $h= - h_0$. Hence $\langle \psi (\tau)
| \psi_G (\tau) \rangle$ will be close to $\langle \psi_G (0)
| \psi_G (\tau) \rangle = 1$. Hence, both $F(\tau)$ and $Q (\tau)$
will be close to zero, as we see in Fig.~\ref{figr3}.

\begin{figure}
\includegraphics[width=0.48\linewidth]{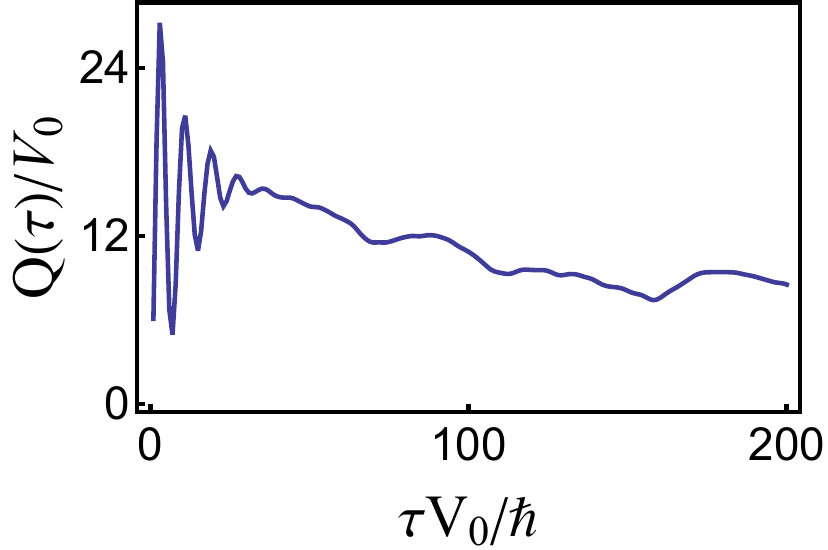}
\includegraphics[width=0.48\linewidth]{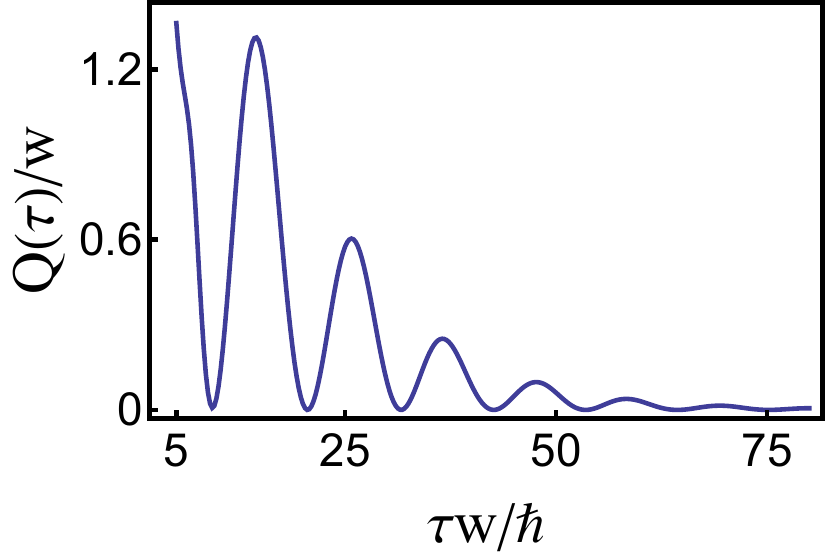}
\caption{Left panel: Plot of $Q(\tau)/V_0$ as a function of $\tau$ for $H$. The ramp 
parameters are same as in Fig.\ \ref{figr3}. Right panel: Similar ramp (Eq.\ \eqref{studr2}) through the critical point of $H'$. See text for details.}
\label{figr4}
\end{figure}

An interesting feature of such a ramp with double passages through an extensively 
degenerate point which distinguishes it from its counterpart in a quantum critical system is the absence of 
St\"uckelberg oscillations. In the latter case, the return probability is known to exhibit St\"uckelberg oscillations \cite{rev4,bhaskar1}. Such an oscillation, for a 
two-state system, can be understood to be due to a relative phase between the amplitudes of the wave functions in the two states \cite{rev4}; a similar phenomenon occurs when a system 
is ramped multiple times through a quantum critical point \cite{bhaskar1}. In contrast, multiple passages through an extensively degenerate point leads to a severe dampening of the 
St\"uckelberg oscillations. This feature can be understood as an effective decoherence due to the presence of an extensive number of relative phases (in contrast to a few relative 
phases for critical systems); averaging over many such phases leads to a decay of these oscillations. This feature can be seen in the left panel of Fig.\ \ref{figr4} where the 
absence of such oscillations can be seen; the right panel shows an analogous plot for $H'$ where the system is ramped twice through the critical point by appropriate variation of $\lambda(t)$ given by
\begin{eqnarray}
\lambda(t) &=& \lambda_c ~+~ \lambda_0 ~\cos(2 \pi t/\tau),\quad 0\le t\le \tau. \label{studr2}
\end{eqnarray}
In this case, $Q(t)$ shows the expected St\"uckelberg oscillations as a function of $\tau$. Thus our analysis suggests that both single and double ramps through a degenerate 
point show distinct features different from analogous ramps through a quantum critical point. 

\section{Periodic dynamics}
\label{periodd} 

In this section, we study a periodic dynamics of $H$ (Eq.\ \eqref{hamsys}). To this end, we vary the transverse field $h$ as a function of time 
using a square-pulse protocol given by
\begin{eqnarray} 
h(t) &=& - ~h_0 \quad {\rm for}~ \,\, t\le T/2 \nonumber\\
&=& h_0 \quad {\rm for}~ \,\, t> T/2. \label{sqp1} \end{eqnarray}
Here $T=2\pi/\omega_D$ denotes the time period of the drive, $\omega_D$ is the drive frequency, and $h_0$ is the drive amplitude. We can obtain some analytical results 
using Floquet perturbation theory which works when the drive amplitude and frequency are
both much larger than the parameter $V_0$. These are presented in Sec.\ \ref{anares}. They are then compared with exact numerical results obtained using exact diagonalization (ED) on a finite chain in Sec.\ \ref{numres}.

\subsection{Analytical results} 
\label{anares} 

In the regime of large drive amplitude, the behavior of a driven system can be studied
using Floquet perturbation theory \cite{rev7}. To this end, we first note that in this 
regime where $V_0 \ll h_0$, one can treat the interaction term perturbatively. The leading order time-evolution operator within this approximation can thus be written as
\begin{eqnarray}
U_0(t,0) &=& e^{-i h_0 \sum_j \sigma^x_j t/\hbar} \quad {\rm for}~ \,\, t\le T/2, \nonumber\\
&=& e^{-i h_0 \sum_j \sigma^x_j (T-t)/\hbar} \quad {\rm for}~ \,\, t> T/2. \label{evol0}
\end{eqnarray} 
Since $U_0(T,0)= I$, we find that $H_F^{(0)}=0$. 

To compute the effect of the interaction term, we use first order perturbation theory which yields
\begin{eqnarray}
U_1(T,0) &=& \left(\frac{-i}{\hbar}\right) \int_0^T U_0^{\dagger}(t,0) H_1 U_0(t,0) \nonumber\\
&=& \left(\frac{-i}{\hbar} \right)\Big( \int_0^{T/2} e^{i h_0 \sum_j \sigma^x_j t/\hbar} H_1 e^{-i h_0 \sum_j \sigma^x_j t/\hbar} \nonumber \\
&& + \int_{T/2}^T e^{i h_0 \sum_j \sigma^x_j (T-t)/\hbar} H_1 e^{-i h_0 \sum_j \sigma^x_j (T-t)/\hbar}\Big). \nonumber \\
&& \label{firstord1}
\end{eqnarray} 
An explicit computation of $U_1$ is most easily done by expressing $U_0$ in the eigenbasis of $\sigma_j^x$ at each site. To this end, we denote
\begin{eqnarray} 
\sigma_j^x |\pm_j\rangle = \pm |\pm_j \rangle, \quad |\pm_j\rangle = \frac{1}{\sqrt{2}} \left( \begin{array}{c} 1 \\ \pm 1 \end{array} \right).
\label{xeiegen1}
\end{eqnarray}
In terms of these states, the standard Pauli matrix operators can be written as 
\begin{eqnarray} 
|s_j\rangle\langle s_j| &=& \frac{1}{2} (1 ~\pm~ \sigma_j^x), \nonumber\\
|s_j\rangle \langle {\bar s_j}| &=& \frac{1}{2} (\sigma_j^z ~-~ {\rm Sgn}[{\bar s_j}] 
~i \sigma_j^y), \label{opmat1}
\end{eqnarray} 
where ${\bar s}_j=\mp$ for $s_j=\pm$, and ${\rm Sgn}[x]$ is the sign function which takes value $1 ~(-1)$ for $x> ~(<)0$. In this basis we can express $U_0(t,0)$ as 
\begin{eqnarray} 
U_0(t,0) &=& \prod_j e^{-i h_0 s_j t/\hbar} |s_j\rangle \langle s_j| \quad {\rm for}~ \,\, 0 \le t\le T/2 \nonumber\\
&=& \prod_j e^{-i h_0 s_j (T-t)/\hbar} |s_j\rangle \langle s_j| \quad {\rm for}~ \,\, 
T/2 \le t \le T. \non \\
&& \label{evol1}
\end{eqnarray} 
Using the fact that $|\uparrow_j\rangle = \sum_{s_j=\pm} |s_j\rangle/\sqrt{2}$, the interaction term $H_1$ can be expressed in this basis as 
\begin{eqnarray} 
H_1 &=& V_0 \sum_j P_j^+ P_{j+1}, ~~~\, P_j = \frac{1}{2} (\sum_{s_j=\pm} |s_j\rangle)(\sum_{s_k=\pm} \langle s_k|) \nonumber\\ \label{hirep} 
\end{eqnarray} 
Note that $P_j$ denotes the projector to $|\uparrow_j\rangle$ and is therefore identical to $(1+\sigma_j^z)/2$. 
Using Eqs.\ \eqref{evol1} and \eqref{hirep}, one can obtain an expression of $U_1$ after a straightforward but somewhat lengthy 
computation detailed in the Appendix. Using $U_1(T,0)= -i H_F^{(1)} T/\hbar$, one finally obtains 
\begin{eqnarray} 
H_F^{(1)} &=& \frac{V_0}{4} \sum_j \Big[ (1+ \frac{1}{2} (\sigma_j^z \sigma_{j+1}^z + \sigma_j^y \sigma_{j+1}^y)) \nonumber \\
&& ~~~~~+ ~\frac{\sin y}{4y} (e^{i y} \sigma_j^+ \sigma_{j+1}^+ + e^{-iy} 
\sigma_j^- \sigma_{j+1}^-) \non \\
&& ~~~~~+ ~\frac{2 \sin (y/2)}{y} (e^{i y/2} \sigma_j^+ + e^{-iy/2} \sigma_j^-)
\Big], \label{fl1}
\end{eqnarray}
where $y= h_0T/\hbar$ and $\sigma_j^{\pm}= \sigma_j^z \mp i \sigma_j^y$. We note that at the special drive frequencies $\omega_D= \omega_p^{\ast} = h_0/(p\hbar)$ (for which $y= 2p \pi$), where $p \in Z$, only the terms in the first line 
of Eq.\ \eqref{fl1} survive. At these frequencies, there is an emergent $U(1)$
symmetry in $H_F^{(1)}$, namely, it is invariant under a rotation in the 
$y-z$ plane by an arbitrary angle. This can be trivially seen from the fact that at these special frequencies, $H_F^{(1)}$ is identical to an $XY$ (here $YZ$) 
Hamiltonian up to an irrelevant constant. This symmetry is emergent since it is not a property of $H(t)$ for any $t$; further, it is approximate since higher order terms in the 
Floquet Hamiltonian do not have this symmetry. This is demonstrated in the Appendix, where the third order Floquet Hamiltonian is explicitly computed. Nevertheless, as 
we shall see in the next section, the presence of such a symmetry shapes the dynamics of the driven system up to a very large prethermal timescale in the high drive 
amplitude regime. 

\subsection{Numerical results} 
\label{numres} 

In this section, we study the driven system using exact numerics. To this end, we denote $H_{-(+)}= H_1-(+) H_0(t)$ for $t\le(>) T/2$. We numerically diagonalize $H_{\pm}$ for system sizes $L\le 18$ 
and obtain the corresponding eigenvalues and eigenstates $\epsilon_n^{\pm}$ and $|n^{\pm}\rangle$ which satisfies $H_{\pm} |n^{\pm}\rangle= \epsilon_n^{\pm} |n_{\pm}\rangle$. We can then 
write the exact evolution operator in terms of these eigenstates and eigenvalues as 
\begin{eqnarray} 
U(T,0) &=& \sum_{n^-,m^+} c^{nm}_{+-} e^{-i(\epsilon_n^+ +\epsilon_m^-) T/\hbar} |n^+\rangle \langle m^-|, \label{evolrep}
\end{eqnarray}
where $c_{+-}^{n m} = \langle n^+|m^-\rangle$. We then diagonalize $U$ numerically to obtain its eigenvalues $\lambda_{\alpha}= e^{i \theta_{\alpha}}$ and eigenfunctions $|\alpha\rangle$. The Floquet quasienergies are obtained from $\lambda_{\alpha}$ using $ \epsilon_{\alpha}^F = - \hbar \arccos ({\rm Re}[\lambda_{\alpha}])/T$.

\begin{figure}
\includegraphics[width=0.98\linewidth]{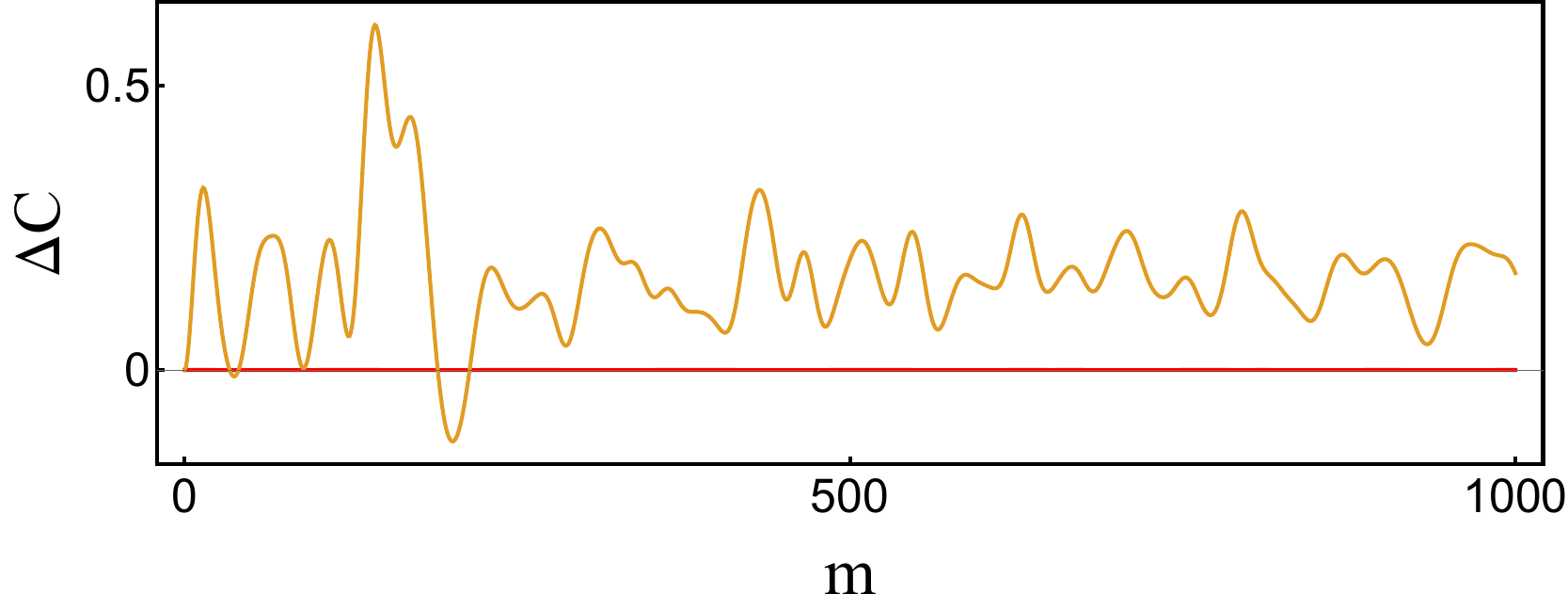}
\includegraphics[width=0.98\linewidth]{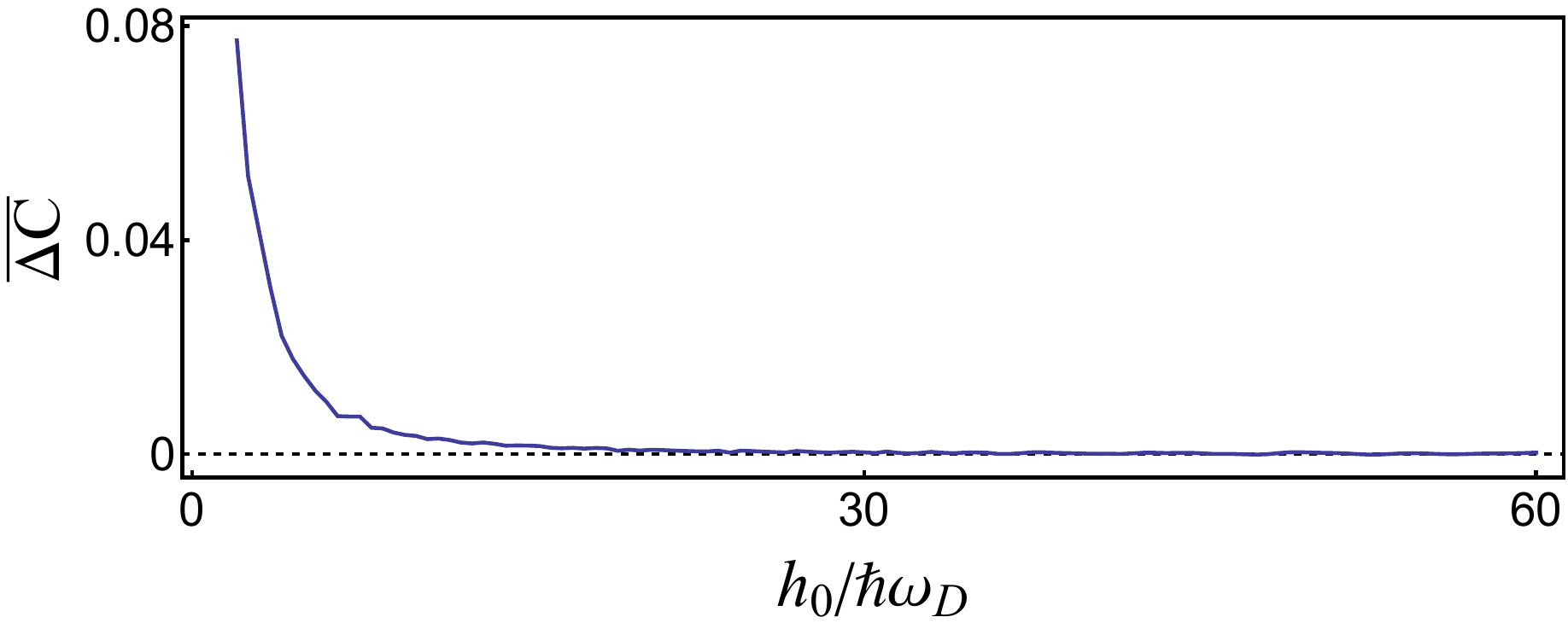}
\caption{Top panel: Plot of $\Delta C(mT)\equiv \Delta C$ as a function of $m$ for
$\omega_D =\omega_1^{\ast}$ (red curve, which stays at zero) and $h_0/(\hbar \omega_D)
= 2/5$ (yellow curve). For both curves, $h_0/V_0= 25$ so that we are in the high drive 
amplitude regime. Bottom panel: Plot of $\overline{\Delta C}$ as a function of $h_0$ 
with $\omega_D=\omega_1^{\ast}= h_0/V_0$. For both figures, $L=16$ and $V_0=1$.}
\label{figfl1}
\end{figure}

To understand the role of the emergent symmetry in the dynamics, we now study the behavior of the correlation functions of the driven chain. To do this, we start from 
an initial state with a product form given by 
\begin{eqnarray}
|\psi_{\rm in}^{(1)}\rangle &=& \prod_j |+_j\rangle, \label{init1}
\end{eqnarray}
and compute the correlation function at stroboscopic times $t=nT$,
\begin{eqnarray}
\Delta C(mT) &=& \langle \psi(mT)| C_{zz} -C_{yy}|\psi(mT) \rangle, \nonumber\\
C_{aa} &=& \sum_j ~\sigma_j^a \sigma_{j+1}^a, \quad a ~=~ z, ~y. \label{corrdef}
\end{eqnarray}

We note that $\Delta C(0)=0$ at $t=0$ for the chosen initial state (Eq.\ \eqref{init1}); 
moreover, it remains zero under evolution by $H_F^{(1)}$ at the special frequencies 
due to the emergent $U(1)$ symmetry which would imply that $| \psi (mT) \rangle = | \psi (0) \rangle$
for all $m$. This is shown in the top panel of Fig.\ \ref{figfl1} where plots of $\Delta C \equiv \Delta C(mT)$ is shown as a function of $m$ for $h_0/V_0=25$. We find 
that in this large drive amplitude regime, where $H_F^{(1)}$ controls the dynamics, $\Delta C$ remains pinned to its initial zero value for spacial drive frequency $\omega_D =\omega_1^{\ast}$ (red curve 
in Fig.\ \ref{figfl1}); in contrast, it shows non-trivial temporal fluctuations when 
driven away from special frequencies ($h_0/(\hbar \omega_D)=2/5$) as shown by the yellow curve in Fig.\ \ref{figfl1}. 
The bottom panel of Fig.\ \ref{figfl1} shows the long-time averaged value of $\Delta C$, given by
\begin{eqnarray}
\overline {\Delta C}= \frac{1}{200} \sum_{m=m_0+1}^{m_0+1000} \Delta C(mT) \label{avc}
\end{eqnarray}
for $m_0=1500$ and in steps of $\Delta m_0=5$ as a function of $h_0/V_0$ with the frequency fixed to $\omega_1^{\ast}$. We find that for large drive amplitudes, where $H_F^{(1)}$ controls the 
dynamics, $\overline {\Delta C}$ remains pinned to zero as expected from the presence of the emergent 
symmetry. In contrast, for lower drive amplitudes, where higher order terms in $H_F$ which do not respect this symmetry become important, $\overline{\Delta C}$ deviates
from zero. This crossover is therefore a signature of the approximate nature of the emergent $U(1)$ symmetry. 

Next, we study the system starting from an initial state given by
\begin{eqnarray} 
|\psi_{\rm in}(\theta)\rangle = \cos \theta |\psi_{\rm in}^{(1)}\rangle + \sin \theta \prod_j |\uparrow_j\rangle \label{init2}
\end{eqnarray}
where $|\uparrow_j\rangle=(1,0)_j^T$ denotes the eigenstate of $\sigma_j^z$ with
eigenvalue $+1$. The initial state is chosen so that it breaks the $U(1)$ symmetry 
exhibited by $H_F^{(1)}$ at $\omega_D=\omega_p^{\ast}$; the degree of this symmetry breaking is parameterized by $\theta$. 

\begin{figure}
\includegraphics[width=0.98\linewidth]{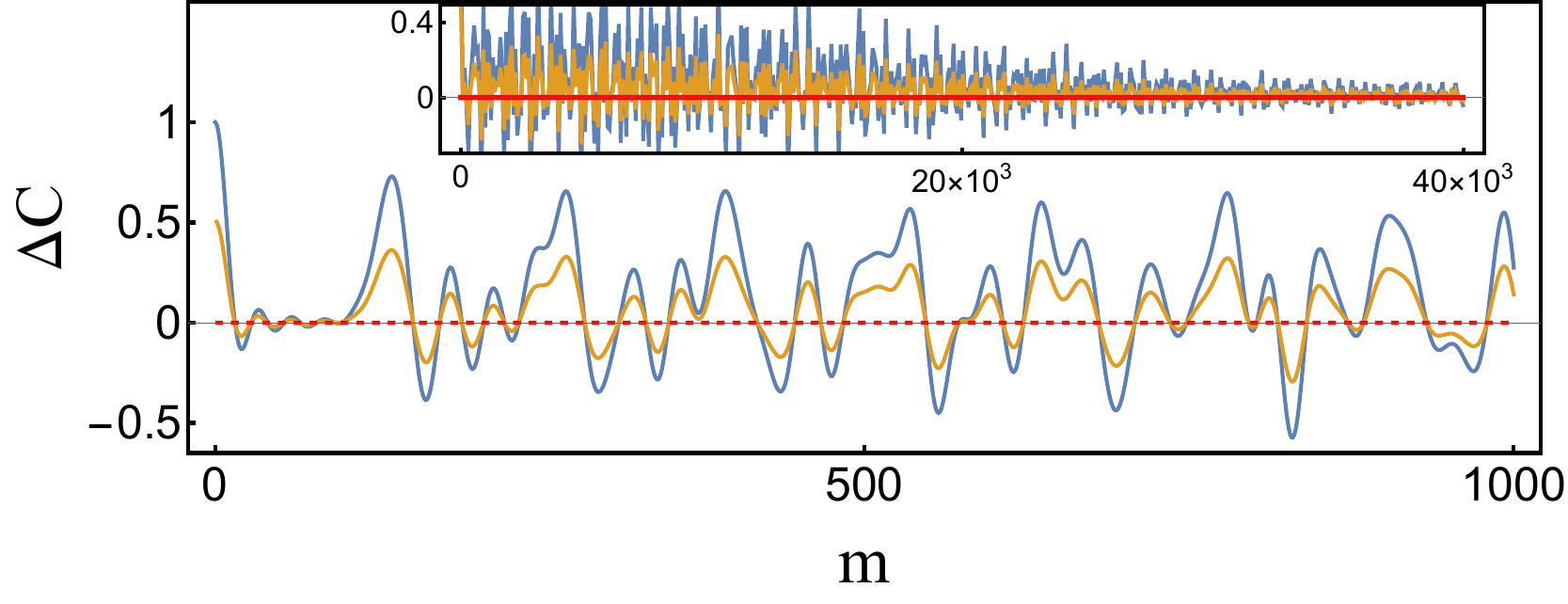}
\includegraphics[width=0.98\linewidth]{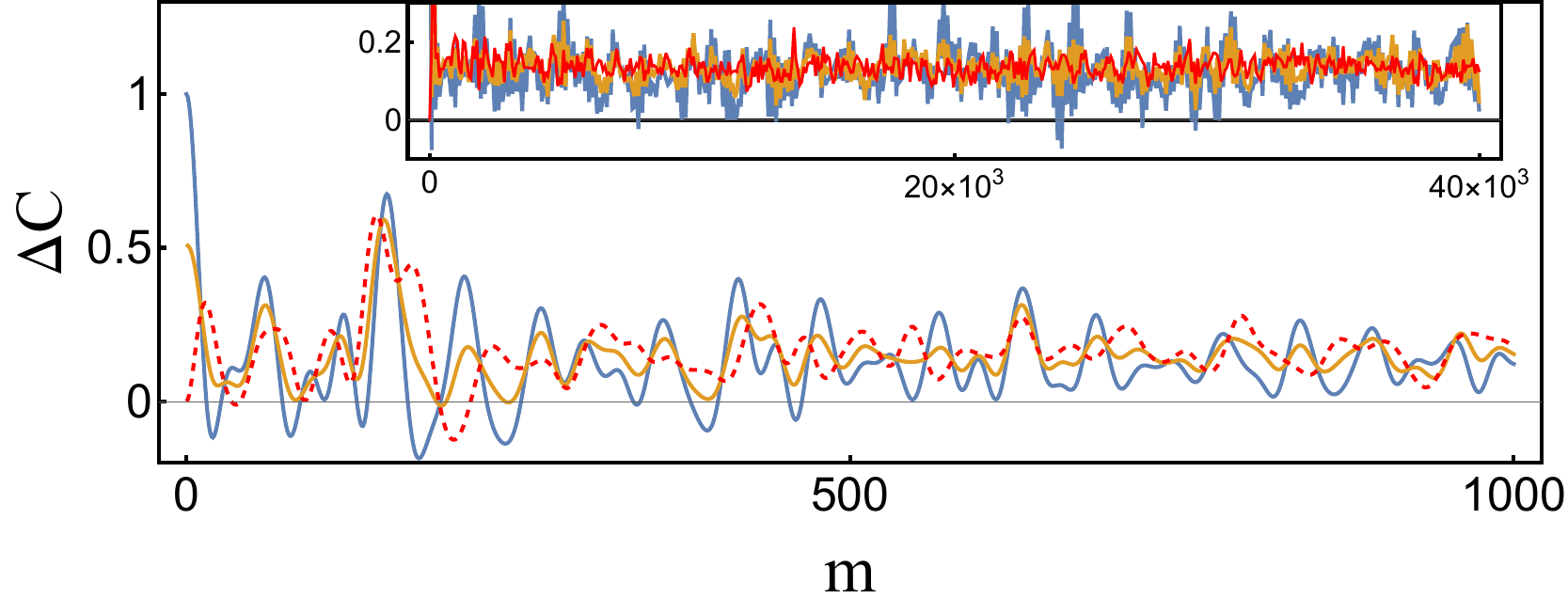}
\caption{Top panel: Plot of $\Delta C(mT)\equiv \Delta C$ as a function of $m$ for
$\omega_D =\omega_1^{\ast}$ and $h_0/V_0 = 25$ for $\theta=0$ (red curve), $\pi/4$ 
(yellow curve) and $\pi/2$ (blue curve). The inset shows the dynamics at a longer timescale. 
Bottom panel: same as the top panel but for $h_0/(\hbar \omega_D)=2/5$. For both figures, $L=16$ and $V_0=1$.
See text for details.} \label{figfl2}
\end{figure}

The dynamics of $\Delta C(mT)$, starting from $|\psi_{\rm in}(\theta)\rangle$, is shown as a function of $m$ in Fig.\ \ref{figfl2}, for several representative values of $\theta$. $\Delta C$ remains pinned to its initial zero value for $\theta=0$ and 
reproduces the behavior shown in Fig.\ \ref{figfl1}. In contrast, for $\theta=\pi/4, \pi/2$, where the initial state breaks $U(1)$ symmetry, it shows non-trivial transient 
dynamics. However, for all the initial states that we have studied, the long-time value of $\Delta C$ vanishes at $\omega_D=\omega_p^{\ast}$. This constitutes dynamic 
restoration of symmetry \cite{cala1,tista1} in the regime where $H_F^{(1)}$ dominates the dynamics. This behavior is to be contrasted with the fate of $\Delta C(mT)$ when $\omega_D \ne \omega_p^{\ast}$; as shown in Fig.\ \ref{figfl2}, $\Delta C(mT)$ remain 
finite for large $m$ away from the special frequency. Thus we expect the dips in the long-time value of $\overline{ \Delta C}$,  starting from an initial symmetry-broken state (Eq.\ \ref{init2}), to provide a 
signature of the dynamic symmetry restoration; we note that this phenomenon has its roots in the emergent symmetry of $H_F^{(1)}$ in the large drive amplitude regime.
 
\section{Discussion}
\label{diss}

To summarize, we have studied the dynamics of a 1D spin model 
which has an exponentially large degeneracy of the ground state for a particular 
value of one of the parameters of the Hamiltonian, namely, when a transverse field
$h=0$. We vary the parameter $h$ in time
using three different protocols: a linear ramp protocol which goes through the degenerate point
$h=0$ once, a cosine ramp protocol for a single cycle which goes through the degenerate point twice in
opposite directions, and a periodic square-pulse protocol where $h$ alternates
between two values which lie on opposite sides of zero. For the ramp protocols, we start with
a state which is the ground state of the initial Hamiltonian while for the periodic drive we start from 
either eigenstate of $H_0$ (Eq.\ \ref{init1}) or superposition of two eigenstates of $H_0$ and $H_1$ (Eq.\ \ref {init2}); 
in each case we evolve the state with the time dependent Hamiltonian.

For the first kind of ramp which traverses the degenerate point once, we find that
both the fidelity $F$ (the log of the overlap between the state at time $t$ and the 
instantaneous ground state at the same time) and the residual energy $Q$ (the 
difference between the expectation value of the Hamiltonian and the instantaneous 
ground state energy) show rapid changes at
the time when the Hamiltonian goes through the degenerate point. Qualitatively,
this occurs because at that time, there is suddenly an exponentially large 
number of states that can easily mix with each other. We compare this
with a different model (the PXP model in a field) which has a conventional quantum critical point where there
are gapless excitations but the ground state is not highly degenerate. In
the latter model, $F$ and $Q$ vary smoothly as we go through the critical point.
In both models, the final values of $F$ and $Q$ are quite
different from their initial values. This is because the final Hamiltonian are quite
different from the initial Hamiltonian. For slow ramps (large ramp time $\tau$),
the dependence of $Q$ on $\tau$ is significantly different for passages through
a degenerate point versus passage through a quantum critical point. Going through a
critical point generates a $Q$ which shows either a Kibble-Zurek scaling 
($Q \sim 1/\tau$) or a Landau-Zener scaling ($Q \sim 1/\tau^2$), with a crossover
between the two regimes depending on the 
ratio between $\tau$ and the system size $L$. 
In contrast, going through a degenerate point generates a $Q$ which either shows
Landau-Zener scaling or a plateau with $\tau$; we never see a Kibble-Zurek scaling.

For the second kind of ramp which goes through the degenerate point twice, both
$F$ and $Q$ change rapidly each time the system goes through the degenerate
point. However, the final and initial values of $F$ and $Q$ are close to each
other because the final and initial Hamiltonian are identical. We find an
interesting difference between the behavior of this model and the PXP model
with a critical point. The PXP model shows prominent oscillations of $Q$ versus
$\tau$; these occur because close to the critical point, the model can be thought 
of as consisting of a 
product to two-state systems, and the relative phases of the amplitudes of the
two states are known to give rise to St\"uckelberg oscillations. In contrast,
the model with extensive degeneracy has such a large number of relative phases that
the St\"uckelberg oscillations are highly damped.

For periodic driving through the degenerate point, we find the interesting phenomenon
that when the drive amplitude is an integer multiple of the drive frequency, and
both are much larger than the other parameters in the Hamiltonian, a $U(1)$
symmetry emerges dynamically. We show this analytically using a Floquet perturbation
theory up to first order. As a consequence of this emergent symmetry, we find 
numerically that as functions of time, certain correlation functions remain close to 
zero (if they are zero to start with) or approach zero if they are not zero in the 
beginning; we call the latter phenomenon as 
dynamic symmetry restoration. The emergent symmetry at the special
frequencies remains valid up to very long times. However, upon reducing the drive amplitude, it breaks down eventually;
we can understand this from Floquet perturbation theory where it is found that
terms beyond the first order do not respect the emergent symmetry.

Experimental verification of our results can be achieved by using a
chain of ultracold Rydberg atoms \cite{rydexp1,rydexp2,rydexp3,rydexp4}. It is well-known
that the effective Hamiltonian for such atoms at zero detuning mimics Eq.\ \eqref{hamsys}. We propose to study such chains in a regime where the Van der Waals interaction between the atoms is 
small enough so that it can be replaced by a nearest-neighbor interaction between the atoms leading to $H_1$ 
in Eq.\ \eqref{hamsys}. The simplest experimental protocol will involve a ramp of $h_0$ with a rate $\tau^{-1}$ for a fixed time $t$; 
such a ramp of the effective 
coupling between the ground state and Rydberg excited states of the atoms through zero can be achieved by a suitable 
manipulation of Raman lasers.  Such a ramp is to be followed by a measurement of the density of Rydberg excitations using standard fluorescence imaging techniques \cite{rydexp1}. 
We expect the average Rydberg excitation density, measured at the end of the ramp, to 
show a variation as a function of the ramp time $t$ analogous to that shown in Fig.\ \ref{figr2}. 

In conclusion, we have studied ramp and periodic dynamics of a quantum Hamiltonian with extensive degeneracy. Our study indicates 
that ramp dynamics of such a Hamiltonian shows a qualitatively different behavior than those of either a gapped system or a critical system. Moreover, the periodic dynamics of 
such a model shows an approximate emergent $U(1)$ symmetry and leads to dynamical symmetry restoration at special drive frequencies. We have suggested experiments which can test our theory.

\vspace{0.6cm}
\centerline{\bf Acknowledgments}
\vspace{0.4cm}

D.S. thanks SERB, India for funding through Project No. JBR/2020/000043.
K.S. thanks DST, India for support through SERB project JCB/2021/000030.

\appendix*

\section {Derivation of Floquet Hamiltonian using Floquet perturbation theory}
\label{app}

In this Appendix we will apply Floquet perturbation theory for 
periodically driven systems \cite{rev11} to our model. (We will set $\hbar =1$ 
here). We consider a time-dependent Hamiltonian of the form
\bea H (t) &=& H_1 (t) ~+~ H_0, \non \\
{\rm where}~~ H_1(t) &=& - ~h(t) ~\sum_j \si_j^x, \label{hamt} \eea
and $H_0$ is time-independent and is given by Eq.\ \eqref{hamsys}. We will consider a square pulse driving protocol 
with
\bea h(t) &=& - ~h_0 ~~{\rm for}~~ 0 ~\le~ t ~<~ T/2, \non \\
&=& + ~h_0 ~~{\rm for}~~ T/2 ~\le~ t ~<~ T, \label{ht} \eea
and $h(t+T) = h (t)$ in general. We will assume that the driving amplitude
and frequency, $h_0$ and $\om = 2 \pi/T$, are both much larger than 
all the coefficients appearing in $H_1$ which is considered to be the
perturbation. We will not make any assumptions about the ratio $h_0/\om$.

Given the form of $H_0 (t)$, the unperturbed time-evolution operator
from time zero to $t$ is given by
\bea U_0 (t,0) &=& {\cal T} e^{-i \int_0^t dt' H_1 (t')} \non \\
&=& e^{-i h_0 t \sum_j \si_j^x} ~~~~~~~~{\rm for}~~ 0 ~\le~ t ~\le~ T/2, \non \\
&=& e^{-i h_0 (T-t) \sum_j \si_j^x} ~~~{\rm for}~~ T/2 ~\le~ t ~\le~ T. \non \\
&& \label{im1} \eea

We now assume a general form of $H_1$ given by
\beq H_0 ~= ~\sum_{m=-\infty}^\infty ~O_m, \label{ham1} \eeq
where $(O_m)^\dagger = O_{-m}$, and
\beq [\sum_j \si_j^x, O_m] ~=~ 2m ~O_m, \label{om} \eeq
Eq.~\eqref{om} implies that in the interaction picture, 
\bea O_m (t) &=& [U_0 (t,0)]^{-1} ~O_m ~U_0 (t,0) \non \\
&=& f_m (t) ~O_m, \non \\
{\rm where} ~~f_m (t) &=& e^{i 2m h_0 t} ~~~~~~~~{\rm for}~~ 0 ~\le~
t ~\le~ T/2, \non \\
&=& e^{i 2m h_0 (T - t)} ~~~{\rm for}~~ T/2 ~\le~ t ~\le~ T. \non \\
\eea
This leads us to define a quantity
\bea I_m &=& \int_0^T dt ~f_m (t) \non \\
&=& -~ \frac{i}{m h_0} ~(e^{i m h_0 T} ~-~ 1) ~~~{\rm if}~~~ m \ne 0,
\non \\
&=& T ~~~{\rm if}~~~ m = 0. \label{im2} \eea

The complete Floquet operator for one time period given by
\beq U (T,0) ~=~ {\cal T} e^{-i \int_0^T dt H (t)} \label{ut1} \eeq
has a perturbative expansion of the form
\beq U(T,0) ~=~ I ~+~ U_1 ~+~ U_2 ~+~ U_3 ~+~ \cdots, \eeq
and the Floquet Hamiltonian $H_F$ defined by $U (T,0) = e^{-i H_F T}$ has a perturbative expansion of the form
\beq H_F ~=~ H_F^{(1)} ~+~ H_F^{(2)} ~+~ H_F^{(3)} ~+~ 
\cdots, \label{hf} \eeq
where
\bea H_F^{(1)} &=& \frac{i}{T} ~U_1, \non \\
H_F^{(2)} &=& \frac{i}{T} ~[U_2 ~-~ \frac{1}{2} (U_1)^2], \non \\
H_F^{(3)} &=& \frac{i}{T} ~[U_3 ~-~ U_1 ~U_2 ~+~ \frac{1}{3} (U_1)^3].
\label{hfx} \eea

Following Ref.\ \onlinecite{rev11}, we find that
\bea U_1 &=& -i ~\sum_m ~\int_0^T dt ~f_m(t) ~O_m \non \\
&=& -i ~\sum_m I_m ~O_m, \label{u1} \eea
and
\beq H_F^{(1)} ~=~ \frac{i}{T} ~U_1 ~=~ \frac{1}{T} ~\sum_m I_m ~O_m. \label{hf1} \eeq

Next,
\bea U_2 &=& (-i)^2 ~\sum_{mn} ~\int_0^T dt_1
\int_0^{t_1} dt_2 \non \\
&& ~~~~~~~~~~~~~~~ f_m (t_1)~ f_n (t_2) ~O_m O_n \non \\
&=& ( -i )^2 ~\sum_{mn} ~c_{mn} O_m O_n, \label{u2} \\
{\rm where}~~~ c_{mn} &=& \int_0^T dt_1 \int_0^{t_1} dt_2 ~f_m (t_1)~
f_n (t_2), \label{cmn} \eea
so that $0 \le t_2 \le t_1 \le T$.

We now prove that $c_{mn} = c_{nm}$. We observe that
\beq f_m (t_1) ~=~ f_m (T - t_1) \eeq
for all values of $t_1$, and similarly for $f_n (t_2)$.
On the right hand side of Eq.~\eqref{cmn}, we re-define variables as
\bea t'_1 = T ~-~ t_2, \non \\
t'_2 = T ~-~ t_1, \eea
so that $0 \le t'_2 \le t'_1 \le T$. Then we get
\beq \int_0^T dt'_1 \int_0^{t'_1} dt'_2 ~f_k (t'_1)~ f_n (t'_2)~, \eeq
which is precisely $c_{nm}$ according to Eq.~\eqref{cmn}. Hence
\beq c_{mn} ~=~ c_{nm}. \label{cmn2} \eeq

We can prove another identity. From Eq.~\eqref{cmn}, we find that
\beq c_{mn} ~+~ c_{nm} ~=~ \int_0^T dt_1 \int_0^T dt_2 ~f_m (t_1)~ f_n (t_2),
\eeq
where there is now no ordering of $t_1, t_2$ on the right hand side;
so we just have two separate integrals.
Doing the integrals explicitly and using Eq.~\eqref{cmn2}, we obtain
\beq c_{mn} ~+~ c_{nm} ~= ~I_m I_n. \label{cmn3} \eeq
Hence
\beq c_{mn} ~=~ c_{nm} ~=~ \frac{1}{2} ~I_m I_n. \label{cmn4} \eeq

Combining Eqs.~\eqref{u1}, \eqref{u2} and \eqref{cmn4}, we see
that
\beq U_2 ~-~ \frac{1}{2} U_1^2 ~=~ 0. \label{hf2} \eeq
Hence $H_F^{(2)} = 0$.

We now turn to
\bea U_3 &=& (-i)^3 ~\sum_{mnk} c_{mnk} ~O_m O_n
O_k, \label{u3} \\
c_{mnk} &=& \int_0^T dt_1 \int_0^{t_1} dt_2 \int_0^{t_2} dt_3~
f_m (t_1) ~f_n (t_2) ~f_k (t_3). \non \\
&& \label{cmnk} \eea
Using arguments similar to the ones which led to Eqs.~\eqref{cmn2} and
\eqref{cmn4}, we obtain
\beq c_{mnk} ~=~ c_{knm} \label{cmnk2} \eeq
(hence only three of the six coefficients are independent, and we will
choose these to $c_{mnk}$, $c_{nkm}$ and $c_{kmn}$), and
\bea && c_{mnk} + c_{mkn} + c_{nmk} + c_{nkm} + c_{kmn} + c_{knm} \non \\
&& = \int_0^T dt_1 \int_0^T dt_2 \int_0^T dt_3 ~f_m (t_1) ~f_n (t_2)~
f_k (t_3), \non \\
&& \eea
where there is now no ordering of $t_1, t_2, t_3$ on the right hand side;
so we just have three separate integrals. Doing these integrals and
using Eq.~\eqref{cmnk2}, we obtain
\beq c_{mnk} ~+~ c_{kmn} ~+~ c_{nkm} ~=~ \frac{1}{2} ~I_m I_n I_k.
\label{cmnk3} \eeq
Using Eq.~\eqref{hf2}, we now proceed to calculate
\bea && U_3 ~-~ U_1 U_2 ~+~ \frac{1}{3} (U_1)^3 ~=~ 
U_3 ~-~ \frac{1}{6} ~U_1^3 \non \\
&& = ~(-i)^3 ~\left[ \sum_{mnk} c_{mnk} ~O_m O_n
O_k ~-~ \frac{1}{6} ~(\sum_m ~I_m O_m)^3 \right]. \non \\
&& \label{u32} \eea

We now have to calculate the coefficient of $O_m O_n O_k$ in Eq.~\eqref{u32}.
We consider three different cases:
(i) $m=n=k$, (ii) $m=n$ but $m \ne k$, and (iii) $m$, $n$ and $k$ are all
different from each other.

\noi Case (i): For $m=n=k$, we find that the terms inside the brackets $\left[
\cdots \right]$ in Eq.~\eqref{u32} give $c_{mmm} O_m^3 - (1/6) (I_m O_m)^3$.
But Eq.~\eqref{cmnk3} gives $c_{mmm} ~=~ (1/6) I_m^3$. Hence the contribution
of $m=n=k$ to $U_3 - (1/6) U_1^3$ is zero.

\noi Case (ii): Suppose that $m=n \ne k$. Then the terms in
Eq.~\eqref{u32} gives
\bea && c_{mmk} ~(O_m O_m O_k + O_k O_m O_m) ~+~ c_{mkm} ~O_m O_k O_m \non \\
&& -~ \frac{1}{6} ~(I_m I_m I_k O_m O_m O_k ~+~ I_m I_k I_m ~O_m O_k O_m 
\non \\
&& ~~~~~~~+~ I_k I_m I_m ~_k O_m O_m). \label{cmmk} \eea
We now use Eq.~\eqref{cmnk3} which gives
\beq I_m I_m I_k ~=~ 4 c_{mmk} ~+~ 2 c_{mkm}. \eeq
Substituting this in Eq.~\eqref{cmmk} and doing some algebra, we get
\beq \frac{1}{3} ~(c_{mmk} ~-~ c_{mkm}) ~[[O_k , O_m], O_m]. \eeq
Adding up all terms of this kind, we get
\beq \frac{1}{3} \sum_{m \ne k} ~(c_{mmk} ~-~ c_{mkm}) ~[[O_k , O_m], O_m].
\label{cmmk2}
\eeq

\noi Case (iii): When $m$, $n$, $k$ are all different, the terms in
Eq.~\eqref{u32} give
\bea && c_{mnk} ~(O_m O_n O_k + O_k O_n O_m) \non \\
&& +~ c_{kmn} ~(O_k O_m O_n + O_n O_m O_k) \non \\
&& +~ c_{nkm} ~(O_n O_k O_m + O_m O_k O_n) \\
&& -~ \frac{1}{6} ~I_m I_n I_k ~(O_m O_n O_k + O_k O_n O_m + O_k O_m O_n 
\non \\
&& ~~~~~~~~~~+~ O_n O_m O_k + O_n O_k O_m + O_m O_k O_n), \non \eea
where we have used Eq.~\eqref{cmnk2}. We now use Eq.~\eqref{cmnk3}.
Collecting all the terms, we get
\bea && c_{mnk} ~[ O_m O_n O_k + O_k O_n O_m \non \\
&& ~~~~~~~~-~ \frac{1}{3} (O_m O_n O_k + O_m O_k O_n + O_n O_k O_m \non \\
&& ~~~~~~~~~~~~~~+ O_n O_m O_k + O_k O_m O_n + O_k O_n O_m) ], \non \\
&& +~ c_{nkm} ~[ ~\cdots~ ] ~+~ c_{kmn} ~[ ~\cdots~ ], \label{cmnk4} \eea
where the $\cdots$ in the last line of Eq.~\eqref{cmnk4} denote terms
obtained from the first three lines of that equation
by doing cyclic permutations of $mnk \to nkm$ and $mnk \to kmn$.
After some algebra, we get
\bea && \frac{1}{3} ~c_{mnk} ~ ( 2 O_m O_n O_k + 2 O_k O_n O_m - O_m O_k O_n \non \\
&& ~~~~~~~~~~~~~- O_n O_k O_m - O_n O_m O_k - O_k O_m O_n) \non \\
&& + ~\frac{1}{3} ~c_{kmn} ~[ ~\cdots~ ] ~+~ \frac{1}{3} ~c_{nkm} ~[ ~\cdots~ ] \non \\
&& = ~\frac{1}{3} ~c_{mnk} ~( [ [O_m, O_n], O_k] ~+~ [ [O_k, O_n], O_m])
\non \\
&& ~+~ \frac{1}{3} ~c_{kmn} ~[ ~\cdots~ ] ~+~ \frac{1}{3} ~c_{nkm} ~[ ~\cdots~ ]. \eea
Adding up all terms of this kind, we obtain
\beq \frac{1}{3} \sum_{m \ne n \ne k} c_{mnk} ~[[ O_m, O_n], O_k], \label{cpqr} \eeq
where the sum runs over all possible cases where $m$, $n$ and $k$ are
different.

Combining Eqs.~\eqref{cmmk2} and \eqref{cpqr}, we get
\bea H_F^{(3)} &=& \frac{i}{T} ~(U_3 ~-~ \frac{1}{6} U_1^3) \non \\
&=& - ~\frac{1}{3T}~ [
\sum_{m \ne k} (c_{mmk} ~-~ c_{mkm}) ~[[O_k , O_m], O_m] \non \\
&& ~~~~~~~~+
\sum_{m \ne n \ne k} c_{mnk} ~[[ O_m, O_n], O_k] ] \non \\
&=& -~ \frac{1}{3T}~ [\sum_{m \ne k} ~(c_{mmk} ~-~ c_{mkm}) ~[[O_k , O_m],
O_m] \non \\
&& ~~~~~~~~~~+ \sum_{m < n < k} \{ (c_{mnk} - c_{kmn}) ~[[ O_m, O_n],
O_k] \non \\
&& ~~~~~~~~~~~~~~~~~+~ (c_{nkm} - c_{mnk}) ~[[ O_n, O_k], O_m] \non \\
&& ~~~~~~~~~~~~~~~~~+~ (c_{kmn} - c_{nkm}) ~[[ O_k, O_m], O_n] \}]. \non \\
\label{hf3} \eea

We now apply the above results to our model in which 
\beq H_1 ~=~ \frac{V_0}{4} \sum_j (1+\sigma_j^z)(1+\sigma_{j+1}^z). 
\label{ham2} \eeq
This can be written in the form shown in Eq.~\eqref{ham1} where only
the following operators appear:
\bea O_0 &=& \frac{V_0}{4} \sum_j ~[1 ~+~ \frac{1}{2} ~( \si_j^z \si_{j+1}^z
~+~ \si_j^y \si_{j+1}^y)], \non \\
O_1 &=& \frac{V_0}{4} \sum_j ~(\si_j^z ~-~ i \si_j^y), \non \\
O_{-1} &=& \frac{V_0}{4} \sum_j ~(\si_j^z ~+~ i \si_j^y), \non \\
O_2 &=& \frac{V_0}{4} \sum_j ~\frac{1}{4} ~(\si_j^z \si_{j+1}^z
~-~ \si_j^y \si_{j+1}^y \non \\
&& ~~~~~~~~~~~~~~~-~ i \si_j^z \si_{j+1}^y ~-~ i \si_j^y \si_{j+1}^z), \non \\
O_{-2} &=& \frac{V_0}{4} \sum_j ~\frac{1}{4} ~(\si_j^z \si_{j+1}^z
~-~ \si_j^y \si_{j+1}^y \non \\
&& ~~~~~~~~~~~~~~~+~ i \si_j^z \si_{j+1}^y ~+~ i \si_j^y \si_{j+1}^z).
\label{o123} \eea

Using Eqs.~\eqref{im2}, \eqref{hf1} and \eqref{o123} , we obtain
\bea H_F^{(1)} 
&=& \frac{V_0}{4} \sum_j ~[~ (1 ~+~ \frac{1}{2} ~( \si_j^z \si_{j+1}^z
~+~ \si_j^y \si_{j+1}^y )) \non \\
&& +~ \frac{2 \sin (h_0 T)}{h_0 T} ~\si_j^z~ -~ \frac{2 (\cos (h_0 T) - 1)}{h_0 T} ~\si_j^y \non \\
&& +~ \frac{\sin (2 h_0 T)}{4 h_0 T} ~(\si_j^z 
\si_{j+1}^z ~-~ \si_j^y \si_{j+1}^y) \non \\
&& -~ \frac{(\cos (2h_0 T) - 1)}{h_0 T}~ (\si_j^z \si_{j+1}^y 
~+~ \si_j^y \si_{j+1}^z)]. \non \\
\label{hf11} \eea
Writing $\sigma^{\pm}_j= \sigma_j^z \mp i \sigma_j^y$, we find that Eq.\ \eqref{hf11} reduces to Eq.~\eqref{fl1} in the main text.

We now note that if $h_0 T = 2 \pi p$, i.e., if
\beq h_0 ~=~ p \om, \label{freeze} \eeq
where $p$ is an integer, we obtain
\beq H_F^{(1)} ~=~ \frac{V_0}{4} \sum_j ~[1 ~+~ \frac{1}{2} ~( \si_j^z \si_{j+1}^z
~+~ \si_j^y \si_{j+1}^y )], \eeq
which commutes with the driving operator $\sum_j \si_j^x$. Hence, to first 
order in Floquet perturbation theory, $\sum_j \si_j^x$ emerges as a
conserved quantity whenever Eq.~\eqref{freeze} is satisfied. 

Next, we consider $H_F^{(3)}$ which can be obtained by substituting Eq.\ \eqref{o123} in Eq.\ \eqref{hf3}. 
We observe that the commutator structure of 
$H_F^{(3)}$ as seen in Eq.\ \eqref{hf3} along with the short-range nature of the 
operators $O_m$ as seen in Eq.\ \eqref{o123} ensure that $H_F^{(3)}$ represents 
a local Hamiltonian. This shows that 
the Floquet Hamiltonian in the perturbative regime has a local structure.

Although the complete expression for $H_F^{(3)}$ in Eq.\ \eqref{hf3} (after substituting
Eq.~\eqref{o123}) is rather cumbersome, we can show that it does not
commute with $\sum_j \si_j^x$ at the special frequencies given in Eq.~\eqref{freeze}.
We first note that the terms $[[O_m, O_n], O_k]$ in Eq.~\eqref{hf3} will commute 
with $\sum_j \si_j^x$ if and only if $m+n+k=0$. We therefore need to look at terms for
which $m+n+k \ne 0$. We discover that the coefficient $c_{001} - c_{010}$ of the term involving $[[O_1, O_0], O_0]$ does not vanish when Eq.~\eqref{freeze}
is satisfied. To be explicit,
\beq c_{001} ~=~ c_{100} ~=~ \frac{T}{4h_0^2} ~+~ \frac{iT^2}{8h_0}, ~~~
c_{010} ~=~ -2 c_{001}, \eeq
in agreement with Eqs.~\eqref{cmnk2} and \eqref{cmnk3} since $I_1 = 0$.
The contribution of this term to $H_F^{(3)}$ is then found to be
\bea && H_F^{(3)} (m=n=0,k=1) \non \\
&& =~ - ~\left( \frac{1}{4h_0^2} ~+~ \frac{iT}{8h_0} \right)~ \left(\frac{V_0}{4}
\right)^3 \non \\
&& ~~~~\times ~\sum_j ~[ i \sigma_{j-1}^z \sigma_{j}^z \sigma_{j+1}^y ~+~
i \sigma_{j-1}^y \sigma_{j}^z \sigma_{j+1}^z \non \\
&& ~~~~~~~~~~ - ~2i \sigma_{j-1}^z \sigma_{j}^y 
\sigma_{j+1}^z ~-~ \sigma_{j-1}^z \sigma_{j}^y \sigma_{j+1}^y \non \\
&& ~~~~~~~~~~ - ~\sigma_{j-1}^y \sigma_{j}^y \sigma_{j+1}^z ~+~ 2 \sigma_{j-1}^y 
\sigma_{j}^z \sigma_{j+1}^y \non \\
&& ~~~~~~~~~~+~ \sigma_{j}^x ~(\sigma_{j-1}^z ~+~ \sigma_{j+1}^z ~-~ i 
\sigma_{j-1}^y ~-~ i \sigma_{j+1}^y) ]. \non \\
&& \label{h001} \eea
The term $H_F^{(3)} (m=n=0,k=-1)$ gives the Hermitian conjugate of the 
expression in Eq.~\eqref{h001}. The other terms of $H_F^{(3)}$ 
can be computed similarly, but we do not list them explicitly here. We also note that 
it can be explicitly checked that $H_F^{(3)} (m=n=0,k=\pm 1)$ do not commute with $\sum_j \sigma_j^x$ 
at the special frequencies. A similar check can be carried out for other terms and this establishes the 
approximate nature of the emergent symmetry.

Before ending, we would like to prove the exact result that $H_F$ is an odd function 
of $V_0$. (This would explain why we found that $H_F^{(2)} = 0$). Given the driving
protocol given in Eqs.~\eqref{hamt} and \eqref{ht}, the Floquet operator in Eq.~\eqref{ut1}
can be written as
\bea U(T,0) &=& e^{-i (T/2) (-h_0 \sum_j \si_j^x ~+~ H_1)} \nonumber\\
&& \times e^{-i (T/2) (h_0 \sum_j \si_j^x ~+~ H_1)}. \label{ut2} \eea
Since all the terms in $H_1$ have the coefficient $V_0$, we denote the operator
in Eq.~\eqref{ut2} as $U(V_0)$. We then observe that
\beq [U(V_0)]^{-1} = U(-V_0). \label{utsym} \eeq
Since $U(V_0) = e^{-i H_F (V_0) T}$, Eq.~\eqref{utsym} implies that 
\beq H_F (-V_0) ~=~ - H_F (- V_0), \eeq
namely, $H_F$ can have only odd powers of $V_0$.


\begin{thebibliography} {99}

\bibitem{frrev1} See, for example, H. L. Stormer, D. C. Tsui, and A. C. Gossard, Rev. Mod. Phys. {\bf 71}, S298 (1999).

\bibitem{slrev1} See, for example, S. Kivelson and S. L. Sondhi, Nature Rev. Phys. {\bf 5}, 368 (2023). 

\bibitem{laughlin1} R. B. Laughlin, Phys. Rev. Lett. {\bf 50}, 1395 (1983).

\bibitem{jain1} J. K. Jain, Phys. Rev. Lett. {\bf 63}, 199 (1989). 
 
\bibitem{slrev2} Y. Zhou, K. Kanoda, and T.-K. Ng, Rev. Mod. Phys. {\bf 89}, 025003 (2017); L. Savary and L. Balents, Rep. Prog. Phys. {\bf 80}, 016502 (2017)

\bibitem{obdref1} J. Villain, R. Bidaux, J. P. Carton, and R. Conte, J. Phys.
{\bf 41}, 1263 (1980); E. F. Shender, JETP {\bf 56}, 178 (1982); 
E. F. Shender and P. C. W. Holdsworth, {\it Order by Disorder and Topology in Frustrated Magnetic Systems}, Pg 259-279 (Springer US, New York, 1996).

\bibitem{obdref2} J. T. Chalker, P. C. W. Holdsworth and E. F. Shender, Phys. Rev. Lett. {\bf 68}, 855 (1992); C. L. Henley, Phys. Rev. Lett. {\bf 62}, 2056 (1989). 

\bibitem{as1}J. M. Hopkinson, S. V. Isakov, H.-Y. Kee, and Y. B. Kim,  Phys. Rev. Lett. {\bf 99}, 037201 (2007);
M. Sarkar, M. Pal, A. Sen, and K. Sengupta, SciPost Phys. {\bf 14}, 004 (2023).
 
\bibitem{rev1} J. Dziarmaga, Adv. Phys. {\bf 59}, 1063 (2010).

\bibitem{rev2} A. Polkovnikov, K. Sengupta, A. Silva, and M. Vengalattore,
Rev. Mod. Phys. {\bf 83}, 863 (2011).

\bibitem{rev3} A. Dutta, G. Aeppli, B. K. Chakrabarti, U. Divakaran,
T. F. Rosenbaum, and D. Sen, {\it Quantum phase transitions in transverse field 
spin models: from statistical physics to quantum information}, (Cambridge University
Press, Cambridge, 2015); {\it Quantum Quenching, Annealing and Computation},
edited by A. Das, A. Chandra, and B. K. Chakrabarti, Lecture Notes in Physics, 
Vol. 802 (Springer, Berlin, Heidelberg, 2010).

\bibitem{rev4} S. N. Shevchenko, S. Ashhab, and F. Nori, Physics
Reports {\bf 492}, 1 (2010). 

\bibitem{rev5} M. Bukov, L. D'Alessio, and A. Polkovnikov, Adv. Phys. {\bf 64}, 139 (2015).

\bibitem{rev6} L. D'Alessio and A. Polkovnikov, Ann. Phys. {\bf 333}, 19 (2013).

\bibitem{rev7} L. D'Alessio, Y. Kafri, A. Polokovnikov, and M. Rigol,
Adv. Phys. {\bf 65}, 239 (2016).

\bibitem{rev8} T. Oka and S. Kitamura, Annu. Rev. Condens. Matter
Phys. {\bf 10}, 387 (2019).

\bibitem{rev9} S. Blanes, F. Casas, J. A. Oteo, and J. Ros, Physics
Reports {\bf 470}, 151 (2009). 

\bibitem{rev10} A. Eckardt, Rev. Mod. Phys. {\bf 89}, 011004 (2017).

\bibitem{rev11} A. Sen, D. Sen, and K. Sengupta, J. Phys. Cond. Mat.
{\bf 33}, 443003 (2021). 

\bibitem{rev12} T. Banerjee and K. Sengupta, J. Phys. Cond. Mat. {\bf 37} 133002 (2025). 

\bibitem{kz1} T. W. B. Kibble, J. Phys. A {\bf 9}, 1387 (1976).

\bibitem{kz2} W. H. Zurek, Nature (London) {\bf 317}, 505 (1985).

\bibitem{ap1} A. Polkovnikov, Phys. Rev. B {\bf 72}, 161201(R) (2005); 
A. Polkovnikov and V. Gritsev, Nature Phys. {\bf 4}, 477 (2008).

\bibitem{ds1} K. Sengupta, S. Mondal, and D. Sen, Phys. Rev. Lett.
{\bf 100}, 077204 (2008).

\bibitem{ds2} D. Sen, S. Mondal, and K. Sengupta, Phys. Rev. Lett.
{\bf 101}, 016806 (2008); R. Barankov and A. Polkovnikov,
Phys. Rev. Lett. {\bf 101}, 076801 (2008).

\bibitem{bhaskar1} B. Mukherjee, P. Mohan, D. Sen, and K. Sengupta, Phys. Rev. B {\bf 97}, 205415 (2018)

\bibitem{dynfr1} A. Das, Phys. Rev. B {\bf 82}, 172402 (2010); S. Bhattacharyya, A. Das, and S. Dasgupta, Phys. Rev. B {\bf 86}, 054410 (2012); S. S. Hegde, H. 
Katiyar, T. S. Mahesh, and A. Das, Phys. Rev. B {\bf 90}, 174407 (2014).

\bibitem{dynfr2} S. Mondal, D. Pekker, and K. Sengupta, Europhys.
Lett. {\bf 100}, 60007 (2012); U Divakaran and K. Sengupta. Phys. Rev. B {\bf 90},
184303 (2014); S. Kar, B. Mukherjee, and K. Sengupta, Phys. Rev. B {\bf 94}, 
075130 (2016); S. Kar, Phys. Rev. B {\bf 95}, 085141 (2017).

\bibitem{dynfr3} H. Guo, R. Mukherjee, and D. Chowdhury,
arXiv:2405.01627 (unpublished); A. Haldar, D. Sen, R. Moessner, and A. Das, Phys.
Rev. X {\bf 11}, 021008 (2021).

\bibitem{dynfr4} G. Camilo and D. Texiera, Phys. Rev. B {\bf 102}, 174304
(2020); B. Mukherjee, R. Melendrez, M. Szyniszewski, H. J.
Changlani, and A. Pal, Phys. Rev. B {\bf 109}, 064303 (2024).

\bibitem{dloc1} A. Agarwala and D. Sen, Phys. Rev. B {\bf 95}, 014305
(2017); S. Aditya and D. Sen, SciPost Phys. Core {\bf 6}, 083 (2023).

\bibitem{dloc2} T. Nag, S. Roy, A. Dutta, and D. Sen, Phys. Rev. B
{\bf 89}, 165425 (2014); L. Tamang, T. Nag, and T. Biswas, Phys. Rev. B {\bf 104},
174308 (2021).

\bibitem{dloc3} Y. Baum, E. P. L. van Nieuwenburg, and G. Refael,
SciPost Phys. {\bf 5}, 017 (2018);  D. J. Luitz, Y. Bar Lev, and A. Lazarides, SciPost Phys. {\bf 3}, 029 (2017).

\bibitem{dloc4} M. Fava, R. Fazio, and A. Russomanno, Phys. Rev.
B {\bf 101}, 064302 (2020); A. Eckardt, C. Weiss, and M.
Holthaus, Phys. Rev. Lett. {\bf 95}, 260404 (2005).

\bibitem{dloc5} R. Ghosh, B. Mukherjee, and K. Sengupta, Phys. Rev.
B {\bf 102}, 235114 (2020); A. C. Keser, S. Ganeshan, G. Refael, 
and V. Galitski, Phys. Rev. B {\bf 94}, 085120 (2016).

\bibitem{topo1} T. Oka and H. Aoki, Phys. Rev. B {\bf 79}, 081406(R) (2009); 
T. Kitagawa, T. Oka, A. Brataas, L. Fu, and E. Demler, Phys. Rev. B {\bf 84}, 
235108 (2011); A. Kundu, H. A. Fertig, and
B. Seradjeh, Phys. Rev. Lett. {\bf 113}, 236803 (2014).

\bibitem{topo2} T. Kitagawa, E. Berg, M. Rudner, and E. Demler, Phys.
Rev. B {\bf 82}, 235114 (2010); N. H. Lindner, G. Refael, and
V. Galitski, Nature Phys. {\bf 7}, 490 (2011).

\bibitem{topo3} M. Thakurathi, A. A. Patel, D. Sen, and A. Dutta, Phys.
Rev. B {\bf 88}, 155133 (2013); M. Thakurathi, K. Sengupta,
and D. Sen, Phys. Rev. B {\bf 89}, 235434 (2015).

\bibitem{topo4} F. Nathan and M. S. Rudner, New J. Phys. {\bf 17}, 125014
(2015); B. Mukherjee, A. Sen, D. Sen, and K. Sengupta, Phys.
Rev. B {\bf 94}, 155122 (2016); B. Mukherjee, Phys. Rev. B {\bf 98}, 235112 (2018).

\bibitem{qscar1} S. Pai and M. Pretko, Phys. Rev. Lett. {\bf 123}, 136401
(2019); B. Mukherjee, S. Nandy, A. Sen, D. Sen, and K. Sengupta,
Phys. Rev. B {\bf 101}, 245107 (2020).

\bibitem{qscar2} K. Mizuta, K. Takasan, and N. Kawakami, Phys. Rev.
Res {\bf 2}, 033284 (2020); S. Sugiura, T. Kuwahara, and K. Saito, Phys. Rev. 
Research {\bf 3}, L012010 (2021); N. Maskara, A. A. Michalidis, W. W. Ho, D. Bluvstein, S. Choi, M. D. Lukin, and M. Serbyn, Phys. Rev. Lett.
{\bf 127}, 090602 (2021)

\bibitem{qscar3} A. Hudomal, J-Y Desaules, B. Mukherjee, G.-X. Su,
J. C. Halimeh, and Z. Papic, Phys. Rev. B {\bf 106}, 104302
(2022); B. Huang, T.-H. Leung, D. M. Stamper-Kurn, and W.
V. Liu, Phys. Rev. Lett. {\bf 129}, 133001 (2022). 

\bibitem{qscar4} B. Mukherjee, A. Sen, D. Sen, and K. Sengupta, Phys.
Rev. B {\bf 102}, 075123 (2020); {\it ibid}, Phys. Rev. B {\bf 102}, 014301 (2020).

\bibitem{som1} S. Ghosh, I. Paul, and K. Sengupta, Phys. Rev. Lett.
{\bf 130}, 120401 (2023); S. Ghosh, I. Paul, and K. Sengupta, Phys. Rev. B 109,
214304 (2024).

\bibitem{dsf} C. M. Langlett and S. Xu, Phys. Rev. B {\bf 103}, L220304
(2021); L. Zhang, Y. Ke, L. Lin, and C. Lee, Phys. Rev. B {\bf 109}, 184313 (2024).

\bibitem{tc1} N. Y. Yao and C. Nayak, Physics Today {\bf 71}, 40 (2018); D. V. Else, 
C. Monroe, C. Nayak, and N. Y. Yao, Ann. Rev. Cond. Mat. {\bf 11}, 467 (2020).

\bibitem{tc2} M. P. Zaletel, M. Lukin, C. Monroe, C. Nayak, F.
Wilczek, and N. Y. Yao, Rev. Mod. Phys. {\bf 95}, 031001
(2023); V. Khemani, R. Moessner, and S. L. Sondhi,
arXiv:1910.10745 (unpublished); K. Sacha and J. Zakrzewski, 
Rep. Prog. Phys. {\bf 81}, 016401 (2018).

\bibitem{tc3} V. Khemani, A. Lazarides, R. Moessner, and S. L. Sondhi, Phys. Rev. 
Lett. {\bf 116}, 250401 (2016); C. W. von Keyserlingk, V. Khemani, and S. L. Sondhi,
Phys. Rev. B {\bf 94}, 085112 (2016); R. Moessner and
S. L. Sondhi, Nature Phys. {\bf 13}, 424 (2017).

\bibitem{tc4} D. V. Else, B. Bauer, and C. Nayak, Phys. Rev. Lett.
{\bf 117}, 090402 (2016); {\it ibid}, Phys. Rev. X {\bf 7}, 011026
(2017); N. Y. Yao, A. C. Potter, I.-D. Potirniche, and A. Vishwanath, 
Phys. Rev. Lett. {\bf 118}, 030401 (2017).

\bibitem{cala1} F. Ares, S. Murciano, and P. Calabrese, Nature Comm. {\bf 14}, 2036 (2023). 

\bibitem{cala2} L. K. Joshi, J. Franke, A. Rath, F. Ares, S. Murciano, F. Kranzl, R. Blatt, P. Zoller, B. Vermersch, P. Calabrese, C. F. Roos, and M. K. Joshi,
Phys. Rev. Lett. {\bf 133}, 010402 (2024).

\bibitem{tista1} T. Banerjee, S. Das, and K. Sengupta, arXiv:2412.03654 
(unpublised).

\bibitem{rydexp1} J. Simon, W. S. Bakr, R. Ma, M. E. Tai, P. M. Preiss,
and M. Greiner, Nature (London) 472, 307 (2011); W. Bakr, A. Peng, E. Tai, R. Ma, 
J. Simon, J. Gillen, S. Foelling, L. Pollet, and M. Greiner, Science {\bf 329}, 
547 (2010).

\bibitem{rydexp2} H. Bernien, S. Schwartz, A. Keesling, H. Levine, A.
Omran, H. Pichler, S. Choi, A. S. Zibrov, M. Endres, M. Greiner, V. Vuletic, 
and M. D. Lukin, Nature {\bf 551}, 579 (2017).

\bibitem{rydexp3} D. Bluvstein, A. Omran, H. Levine, A. Keesling, G. Semeghini,
S. Ebadi, T. T. Wang, A. A. Michailidis, N. Maskara, W. W. Ho, S. Choi, M. Serbyn,
M. Greiner, V. Vuletic, and M. D. Lukin, Science {\bf 371}, 1355 (2021).

\bibitem{rydexp4} S. Ebadi, T. T. Wang, H. Levine, A. Keesling, G. Semeghini,
A. Omran, D. Bluvstein, R. Samajdar, H.
Pichler, W. W. Ho, S. Choi, S. Sachdev, M. Greiner,
V. Vuletic, and M. D. Lukin, Nature {\bf 595}, 227 (2021) 

\bibitem{subir1} S. Sachdev, K. Sengupta, and S. M. Girvin, Phys. Rev.
B {\bf 66}, 075128 (2002).

\bibitem{subir2} P. Fendley, K. Sengupta, and S. Sachdev, Physical Review
B {\bf 69}, 075106 (2004).

\bibitem{pekker1} M. Kolodrubetz, D. Pekker, B. K. Clark, and K. Sengupta, Phys. Rev. B {\bf 85}, 100505(R) (2012); R. Ghosh, A. Sen, and K. Sengupta, Phys. Rev. 
B {\bf 97}, 014309 (2018).
 
\end{thebibliography}
\end{document}